\documentclass{aa}  
\usepackage{graphicx}
\usepackage{txfonts}
\usepackage{aalongtable}
\usepackage{natbib}
\bibpunct{(}{)}{;}{a}{}{,}

\def\cm-2{cm$^{-2}$}

\def\msun{$M_{\odot}$}

\def\chandra{{\it Chandra}}

\def\xmm{{XMM-Newton}}
\def\m31{M~31}
\def\me33{M~33}

\newcommand{\ergs}[1]{$\times 10^{#1}$ \hbox{erg s$^{-1}$}}

\newcommand{\hcm}[1]{$\times 10^{#1}$ cm$^{-2}$}

\newcommand{\nh}{\hbox{$N_{\rm H}$}}

\begin{document}
   \title{X-ray monitoring of optical novae in M~31 from July 2004 
   to February 2005\thanks{Partly 
   based on observations obtained with the Wendelstein
    Observatory of the Universit\"atssternwarte M\"unchen}
    \fnmsep\thanks{Tables~A.1 and A.2 are 
    only available in electronic form
    at the CDS via anonymous ftp to cdsarc.u-strasbg.fr (130.79.128.5)
    or via http://cdsweb.u-strasbg.fr/cgi-bin/qcat?J/A+A/ }}


   \author{W.~Pietsch\inst{1} \and 
	   F.~Haberl\inst{1} \and
	   G.~Sala\inst{1} \and
	   H.~Stiele\inst{1} \and 
	   K.~Hornoch\inst{2} \and
	   A.~Riffeser\inst{1,3} \and
           J.~Fliri\inst{3} \and
           R.~Bender\inst{1,3} \and
	   S.~B\"uhler\inst{3} \and
           V.~Burwitz\inst{1} \and
	   J.~Greiner\inst{1} \and
	   S.~Seitz\inst{3} 
          }


   \institute{Max-Planck-Institut f\"ur extraterrestrische Physik, Giessenbachstra\ss e, 
           85741 Garching, Germany \\
	   \email{wnp@mpe.mpg.de}
	 \and
	   66431 Lelekovice 393, Czech Republic
         \and
	   Universit\"atssternwarte M\"unchen, Scheinerstra\ss e, 81679
	   M\"unchen, Germany 
            }

   \date{Received 27 November 2006 / Accepted 12 December 2006}

  \abstract
   {Optical novae have recently been identified as the major class of supersoft
   X-ray sources in M~31 based on ROSAT and early \xmm\ and \chandra\ observations.}
   {This paper reports on a search for X-ray counterparts of optical novae in 
   M~31 based on archival \chandra\ HRC-I and ACIS-I as well as \xmm\ 
   observations of the galaxy center region
   obtained from July 2004 to February 2005. }
   {We systematically determine X-ray brightness or upper limit for counterparts
   of all known optical novae with outbursts between
   November 2003 to the end of the X-ray coverage. In addition, we determine 
   the X-ray brightnesses for counterparts of four novae with earlier outbursts.}
   {For comparison with the X-ray data we created a catalogue of optical 
   novae in M~31 based on our own nova search programs and on all novae 
   reported in the literature. We collected all known properties and named the 
   novae consistently following the 
   CBAT scheme.  We detect eleven out of 34 novae within a year after 
   the optical outburst in X-rays. 
   While for eleven novae we detect the end of the supersoft 
   source phase, seven novae are still bright more than 1200, 1600, 1950, 2650,
   3100,
   3370 and 3380~d after outburst. 
   One nova is detected to turn on 
   50~d, another 200~d after outburst. Three novae unexpectedly showed short 
   X-ray outbursts starting
   within 50~d after the optical outburst and lasting only two to three months. The
   X-ray emission of several of the novae can be characterized as
   supersoft from hardness ratios and/or X-ray spectra or by comparing HRC-I
   count rates with ACIS-I count rates or upper limits.}
   {The number of detected optical novae at supersoft X-rays is much higher than previously estimated ($>30\%$). 
   We use the X-ray light curves to estimate the 
   burned masses of the White Dwarf and of the ejecta.}


   \maketitle
%

\section{Introduction}
The outbursts of classical novae (CNe) are caused by the explosive hydrogen burning
on the white dwarf (WD) surface of a cataclysmic variable (CV), a close binary
system with transfer of matter from a main sequence star to the WD.
After $10^{-5}-10^{-4} M_{\odot}$ of H-rich material are transfered to the WD, 
ignition in degenerate conditions takes place in the accreted
envelope and a thermonuclear runaway is initiated \citep{1998ApJ...494..680J,
1998MNRAS.296..502S,1995ApJ...445..789P}.
As a consequence, the envelope expands and causes the
brightness of the star to increase to
maximum luminosities up to $\sim 10^{5} L_{\odot}$. A fraction
of the envelope is ejected, 
while a part of it remains in steady nuclear burning
on the WD surface. This powers a supersoft X-ray source (SSS)
which can be observed as soon as the expanding ejected envelope becomes 
optically thin to soft X-rays, with the spectrum of a hot 
($T_{eff}: 10^{5}-10^{6}$ K) WD atmosphere 
\citep{1991ApJ...373L..51M}.
The duration of the SSS phase is inversely related to the 
WD mass while the turn-on of the SSS is determined by the mass ejected in the
outburst. Models of the post-outburst WD envelope show that  
steady H-burning can only occur for masses smaller than $\sim 10^{-5} M_{\odot}$ 
\citep{2005A&A...439.1061S,1998ApJ...503..381T}, and the observed evolution
of the SSS in V1974 Cyg has been successfully modeled by an envelope of 
$\sim 2\times10^{-6} M_{\odot}$ \citep{2005A&A...439.1057S}. 
WD envelope models also show that the duration of the SSS state depends 
on the metalicity of the envelope, so the monitoring of the SSS states of CNe 
provides constraints also on the chemical composition of the post-outburst envelope. 
\citet{2006ApJS..167...59H} have
developed envelope and optically thick wind models that simulate the optical, UV
and X-ray light curves for several WD masses and chemical compositions, and have
used them to successfully simulate the observed light curves of several novae
\citep{2005ApJ...631.1094H,2005ApJ...633L.117K,2006astro.ph.11594K}. 

Accreting WDs in recurrent novae (RNe) are good candidates for type
Ia supernovae (SNe) as RNe are believed to contain massive WDs.
However, one of the main
drawbacks to make RNe believable progenitors of SNe-Ia was their low fraction
in optical surveys \citep{1994ApJ...423L..31D}. 
In the case of CNe the ejection of
material in the outburst makes it difficult to follow the
long-term evolution of the WD mass. For some CNe there
is a disagreement between theory and observations regarding the ejected
masses, with observational determinations of the mass in the ejected
shell larger than predicted by models. The duration of the SSS state 
provides the only direct 
indicator of the post-outburst envelope mass in RNe and CNe. In the case of
CNe with massive WDs, the SSS state is very short ($<$100~d) and could have been
easily missed in previous surveys. 
CNe with short SSS state are additional good candidates for SNe-Ia progenitors 
which makes determining their frequency very important.

Nevertheless, the number of SSS states observed in novae 
is small. In a total of 39 CNe observed less than ten years after the 
outburst by ROSAT, 
SSS states were found only in three novae \citep{2001A&A...373..542O}, with
SSS phases lasting between 400 days and 9 years: 
GQ Mus \citep{1993Natur.361..331O}, 
V1974 Cyg \citep{1996ApJ...456..788K}, 
and Nova LMC 1995 \citep{1999A&A...344L..13O,2003ApJ...594..435O}.
The \chandra\ and \xmm\ observatories have detected SSS emission for three more 
novae:
V382 Vel \citep{2002MNRAS.333L..11O,2002AIPC..637..377B},
V1494 Aql \citep{2003ApJ...584..448D}, 
and V4743 Sgr \citep{2003ApJ...594L.127N,2004IAUC.8435....2O}.
But for most Galactic and LMC novae, only a limited number of observations 
have been performed for each source, 
providing little constraints on the duration of the SSS state. 
A recent exception is
the monitoring of the recurrent nova RS~Oph in spring 2006 with
the Swift satellite which clearly determined the end of the SSS state after less than
100 days after outburst \citep[see e.g.][]{2006ATel..838....1O} 
which suggests a WD mass of 1.35 M$_{\odot}$ \citep{2006ApJ...651L.141H}.

Although only X-ray observations can provide direct insight into the hot
post-outburst WD, ultraviolet emission lines arising from the
ionization of the ejecta by the central X-ray source reflect the presence
of on-going hydrogen burning on the WD surface. Several works
have used this indirect indicator to determine the turn-off of classical
novae from IUE observations 
\citep{1996ApJ...463L..21S,1998A&AS..129...23G,2001AJ....121.1126V},
showing in all cases turn-off times shorter than expected.

The small number of novae found to exhibit a SSS state,
and the diversity of the duration of this state (from 10 years down
to few weeks) are one of the big mysteries in the study of 
hydrogen burning objects over the last years. Despite an extensive
ToO program with \chandra\ and \xmm\ (of order 3 dozen observations
during the last 6 years), little progress has been made in
constraining the duration of the SSS states, or to even
putting constraints on the long term evolution of accreting WDs in 
binary systems.

In contrast to the Galaxy, X-ray observations of the central area of the big 
nearby spiral galaxy 
\m31\ \citep[distance 780 kpc,][]{1998AJ....115.1916H,1998ApJ...503L.131S},
with its moderate Galactic foreground absorption  \citep[\nh = 6.66\hcm{20},
][]{1992ApJS...79...77S},
offer the unique chance to learn more about the duration of the SSS phase
in novae with minimal effort: \m31\ is the only nearby galaxy with many (more
than 100 nova explosions are known from the center area over the last 5 years!)
reported optical novae within the field of view (FOV) of one \xmm\ EPIC or
\chandra\ HRC-I or ACIS observation. All novae are at the same, known distance,
thus allowing easy comparison of light curves and maximum brightness/luminosity in
optical and X-rays.

Recently, \citet[][hereafter PFF2005]{2006A&A...454..773P,2005A&A...442..879P} combined an optical 
nova catalogue from the WeCAPP survey with optical novae reported in the 
literature and correlated them with the most recent X-ray catalogues
from ROSAT, \xmm\ and \chandra, and -- in addition -- searched for nova 
correlations in archival data. They reported 21 X-ray counterparts for novae 
in \m31\ -- mostly identified as supersoft sources (SSS) by their hardness 
ratios -- and two in \me33. Their sample more than triples the number of known 
optical novae with supersoft X-ray phase. For most of the counterparts, X-ray 
light curves could be determined. From the well determined start times of the 
SSS state in two novae, the hydrogen mass ejected in the outburst could be
determined to $\sim10^{-5} M_{\odot}$ and $\sim10^{-6} M_{\odot}$, respectively.
The supersoft X-ray phase of at least 15\% of the novae started within a year. 
At least one of the novae showed a SSS state lasting 6.1 years after
the optical outburst. Six of the SSSs turned on between 3 and 9 years 
after the optical discovery of the outburst and may be interpreted as 
recurrent novae. If confirmed, the detection of a delayed SSS phase turn-on 
may be used as a new method to classify novae as recurrent. The 
new method yielded a ratio of recurrent novae to classical novae of 0.3 which 
is in agreement (within the errors) with previous works.

For one of these six cases \citep[source 191 from the \xmm\ catalogue of \m31\ 
X-ray 
sources][hereafter PFH2005]{PFH2005}, \citet{2006IBVS.5720....1S} 
showed that the SSS was correlated with a nova which was detected on optical
images 84 days before the X-ray detection (M31N~2001-10f). 
The position of this nova differs
from that of the nova close-by proposed as identification by PFF2005. Therefore 
the explanation as recurrent nova is no longer necessary for this system.

Recently, \citet{2006IBVS.5737....1S} reported the identification of
another transient SSS in the \m31\ northern disk, [PFH2005] 543 = 
XMMU J004414.0+412204 = n1-86 
\citep{2002IAUC.7798....2T,
2002ATel...82....1G,2005ApJ...634..314T,2006ApJ...643..356W}, with an optical
nova (M31N~2001-11a, see Table A.1). The nova outburst in November 2001 
happened 53 days before the detection as SSS. 
After the first SSS in \m31\ identified with a nova, M31N~1990-09a
\citep[][]{2002A&A...389..439N} and M31N~2001-10f,
M31N~2001-11a is already the third 
optical nova in \m31\ for which the optical outburst was
detected when searching for optical counterparts of supersoft X-ray
sources.

Here we report on a follow-up of the PFF2005 work based on archival \chandra\ HRC-I and ACIS-I 
as well as \xmm\ observations of the \m31\ center area collected from July 2004
to February 2005. The
optical nova catalog used for the correlation contains many novae detected by 
Kamil Hornoch (Appendix~B) and the WeCAPP project and additional 
novae from the literature (see Sect.~2 and Appendix~A). 
During this work we created an \m31\ optical nova catalog from the literature
that contains $\sim$700 optical nova candidates with references for parameters
(like positions,
date of outburst, brightness and filter, optical spectra, X-ray detection). We
homogeneously name the \m31\ novae following the CBAT naming scheme (see Appendix~B) that we want to
update regularly in an internet version.
In Sect.~3, we discuss the X-ray observations and methods used.
In Sect.~4 we give light curves and upper limits, and X-ray spectral information 
for counterparts of \m31\ novae.
Based on these results we argue that X-ray SSS states are significantly more
common in optical novae than reported before. We discuss ejected masses and WD
masses derived from the X-ray light curves. 
We also discuss the special case 
of nova M31N~2004-11f, the first nova in \m31\ detected in Hubble Space
Telescope (HST) images and probably a recurrent nova, which is one of the two
novae in \m31\ detected
in X-rays within 35 days after the optical outburst (Sect.~5). 

\section{M~31 optical nova catalogue}\label{sect:opt}
The optical novae used for cross-correlation with the X-ray data
result in part from 8 years of observations (September 1997 to February 2005) 
of the
central part of \m31\ by the continuing Wendelstein Calar Alto
Pixellensing Project (WeCAPP, \citealt{2001A&A...379..362R}).  WeCAPP
monitors a $17.2\arcmin \times 17.2\arcmin$ field centered on the
nucleus with the 0.8~m telescope at Wendelstein Observatory (Germany)
continuously since 1997. The observations are
carried out in $R$ and $I$ filters close to the Kron-Cousins system. Data were
reduced using the WeCAPP reduction pipeline {\sf mupipe}, which
implements an image subtraction technique \citep{1998ApJ...503..325A}
to overcome the crowding effects and allow proper photometry of
variable sources in the central bulge of \m31\
\citep{2003ApJ...599L..17R}.

In the full WeCAPP data set, 23781 variable sources were detected,
most of them being Long Period Variables \citep{2006A&A...445..423F}.
The 1$\sigma$ error radius of the astrometric solution is 0\farcs16.

A catalogue of more than 75 novae brighter than 20.0 mag detected in the survey
is in preparation \citep[][hereafter FBR2007]{fliri2007}. 
The outburst of six of the novae in 2005 occurred
after the X-ray observations. An example for an optical light curve of
a nova which correlates with a time variable SSS detected by \xmm\ and
\chandra, was shown in Fig.~1 of PFF2005 (nova M31N~2000-07a =
WeCAPP-N2000-03).

We combined the WeCAPP nova list with novae from other 
surveys of \m31. Many of them are listed in the nova pages 
``M31 (Apparent) Novae Page"
provided by the International Astronomical Union, Central Bureau for 
Astronomical Telegrams CBAT and the finding
charts and  information, collected by David Bishop (see Appendix A).
Of specific interest for the X-ray correlations were the many novae detected by 
Kamil Hornoch and collaborators in 2003 to 2005 for which we give position and 
light curve information in Appendix B.
Throughout the paper, we use the CBAT nova nomenclature (see Appendix A). We 
also adopted this naming scheme for novae that were not registered by CBAT.
For the WeCAPP and Hornoch novae we estimated the time interval between maximum
brightness and two magnitudes below. We derived these $t_2$ timescales 
with a very simple algorithm.
First we evaluated the observed maximum brightness and its Julian date.
Starting from this point we searched for later points, where the
interpolated (decreasing) line between two data points is crossing
a brightness 2 mag fainter than the maximum. The difference
between this time and the maximum time gives a good guess
for the $t_2$ timescale. 

For the detailed search for X-ray emission, we use all optical novae in the FOV
of the \chandra\ HRC-I with 
outbursts from 13 months before to the time of 
the deep X-ray observations (i.e. five in November/December 2003, 29 in 2004 and
two in January/February 2005).

There are three optical novae in 2004 that need a special discussion:

M31N~2004-07a was detected by Fiaschi et al. (2004, see
http://cfa-www.harvard.edu/iau/CBAT\_M31.html) on H$\alpha$  filter images 
on July 30, 2004. In the WeCAPP catalogue there is a nova within 4\arcsec\ of
the position (WeCAPP~N2004-06) which was detected on R band images on November
5, 2004 when that field was covered for the first time after a gap of nearly
half a year. A search on Ond\v{r}ejov observatory R filter and 
La Palma observatory H$\alpha$ filter images (see 
Nova No.41 in Appendix B) revealed that both detections are from the same nova
outburst, that the position of the nova is close to that given by WeCAPP, and
that it was a slow nova with an outburst well before the first H$\alpha$ filter 
detection.

Nova M31N~2004-11g  was detected in the WeCAPP survey (WeCAPP~N2004-10) 
on November 10, 2004 after a gap in the observations of 15 days. A search on 
Lelekovice and Ond\v{r}ejov R filter images (see Nova No.40 in Appendix B) 
revealed that the nova outburst occurred at least 4 days earlier. 

\begin{table}
\begin{center}
\caption[]{Detection of novae within 2\arcsec\ of M31N~2004-11f.}
\begin{tabular}{lrrrcl}
\hline\noalign{\smallskip}
\hline\noalign{\smallskip}
\multicolumn{1}{l}{Name$^a$} & \multicolumn{1}{c}{RA~~~(h:m:s)}
&\multicolumn{1}{c}{$\Delta_{\rm RA}$} &\multicolumn{1}{c}{JD$^b$}
&\multicolumn{1}{c}{Brightness} &Ref.$^c$\\
M31N&\multicolumn{1}{c}{Dec~~(d:m:s)} 
&\multicolumn{1}{c}{$\Delta_{\rm Dec}$ } 
&\multicolumn{1}{c}{T (a)$^d$} &\multicolumn{1}{c}{(mag) }&\\
& \multicolumn{1}{c}{J2000} & (\arcsec) & & \multicolumn{1}{c}{Band}\\
\noalign{\smallskip}\hline\noalign{\smallskip}
1984-07a & 00:42:47.2 & 0.6 & 5909.5& 17.6:& (1)\\
         &+41:16:20   & 0.2 &  0.0  &  B \\
2001-10c &00:42:47.21 & 0.7 &12193.5 & 17.4 & (2) \\
         &+41:16:18.7  & 1.1 &  17.2 &  R \\
2004-02a &00:42:47.27 & 1.4 &13039.3 & 16.6 & (3) \\
         &+41:16:21.4 & 1.6 &   19.5 &  R \\
2004-11f &00:42:47.15 & 0.0 & 13311.8 & 17.9 & (4) \\
         &+41:16:19.8 & 0.0 &   20.3 &  R \\
\noalign{\smallskip}
\hline
\noalign{\smallskip}
\end{tabular}
\label{tab:rec}
\end{center}
Notes: \\
\hspace{0.3cm} $^a$: following CBAT nomenclature (see text) \\
\hspace{0.3cm} $^b$: Julian Date - 2\,440\,000 \\
\hspace{0.3cm} $^c$: Detection reference: 
(1) R137, \citet{1989AJ.....97...83R}
(2) WeCAPP~N2001-14, FBR2007
(3) WeCAPP~N2004-03, FBR2007 
(4) WeCAPP~N2004-11, FBR2007, this work\\
\hspace{0.3cm} $^d$: Time from first reported outburst in the field in years\\
\end{table}

\begin{figure*}
   \includegraphics[width=18cm,clip]{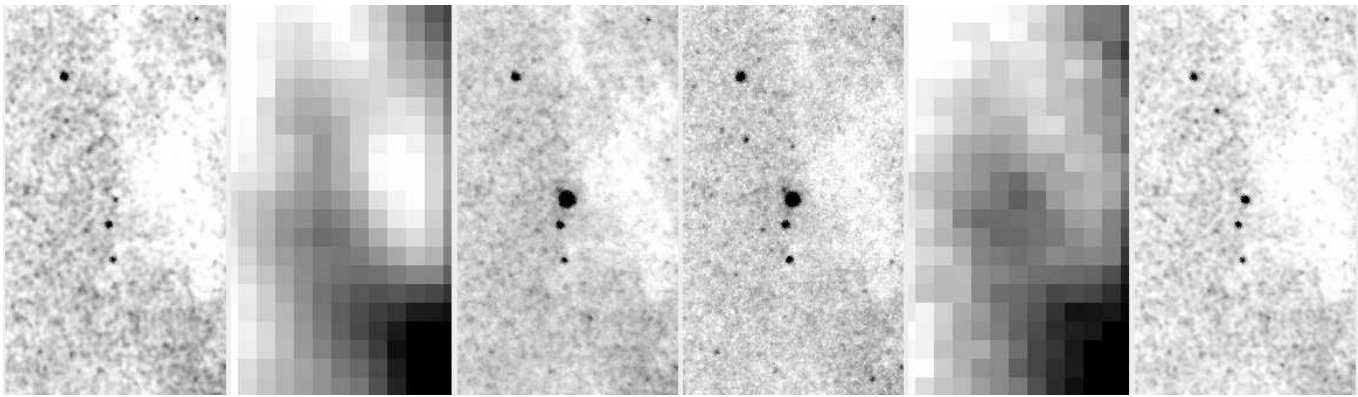}
      \caption{Images of
      M31N~2004-11f:
 From the left to the right  HST-ACS (F435W) 2004-01-23,  Wendelstein 
(R) 04-10-25,  HST-ACS (FR423N) 2004-11-02, HST-ACS (FR388N) 2004-11-08, 
Wendelstein (R) 04-11-09, HST-ACS (F435W) 2004-11-22. The different 
bands are given in brackets.
The angular scale of each image is 6" x 10.5".
Note that already in the first image the 
pre-nova is clearly identified. Due to the lower resolution the faint 
source is not visible in the second image taken on Wendelstein previous 
to the nova outburst. After the outburst the nova is also visible in the 
fifth image from Wendelstein.}
         \label{fig:hst}
\end{figure*}
\begin{figure}
\centering
\includegraphics[width=8.5cm]{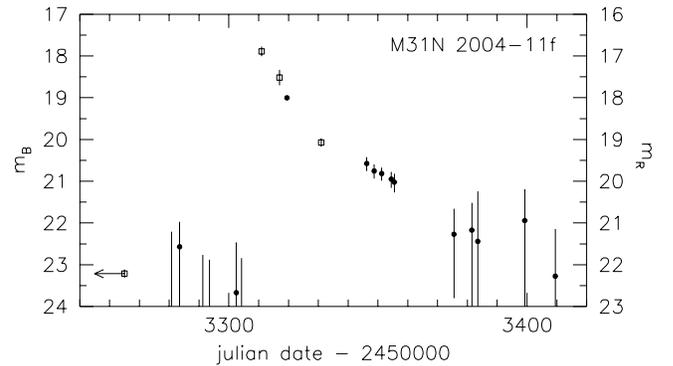}
   \caption{In this figure we present the first nova in \m31\ detected 
   in ground and space data (HST). The
   Nova M31N~2004-11f light curve consists of four data
points obtained with HST/ACS in the blue band using the F435W
(JD=2453027 and JD=2453331), FR423N (JD=2453311) and FR388N
(JD=2453317) filters, respectively and plotted as open squares.
The first HST F435W image provides a good estimate for the brightness about 
9 months 
before the eruption (the arrow-marked data point), which translates in a 
minimum outburst amplitude of 5.3 mag in the B band and a maximum (observed) 
brightness of B = 17.9 mag (see scale at left y-axis). Additionally
 16 WeCAPP data points were obtained in the R-band (scale at right y-axis)
with the 0.8~m Wendelstein telescope
which are shown as filled circles with their 3$\sigma$ error bars. 
	   }
      \label{fig:2004_11f_lc}
\end{figure}

Nova M31N~2004-11f was detected in the WeCAPP survey 
(WeCAPP~N2004-11) on November 10, 2004 in the same field. Its position close to
the \m31\ center prevented the detection in Lelekovice and Ond\v{r}ejov images.
However, a search in archival HST ACS images showed the nova in outburst on
three images on 2004 November 2, 8, and 22 (8 and 2 days before and 12 days
after the WeCAPP detection), while it is also detected in the pre-nova 
stage on 2004 January 23. For the ACS
data we carried out an absolute photometric calibration in the F435W
band for the date JD=2453331 using a zero point in the B band of
25.779 mag. This is consistent with the zero point given in
\citet{2004STScI_08_ISR}. The absolute calibration was checked using
the globular cluster [BHB2000] 124-NB10 yielding 15.94 mag in B
consistent with the published \citep{2000AJ....119..727B} value of 15.87
mag. We thus estimate an accuracy for the F435W-band measurements of
roughly 0.1 mag.  All HST/ACS data points were obtained by differential
photometry in respect to four different stars in each ACS image. The
scatter of the magnitudes derived from these stars give a rough
estimate for the errors (see Figs.~\ref{fig:hst} and \ref{fig:2004_11f_lc}). 
Already in the HST image taken about 9 months before the eruption the 
pre-nova is clearly identified. Its estimated brightness in outburst of 
B = 17.9 mag and 
the minimal outburst amplitude of 5.3 mag in the B band are well consistent 
with M31N~2004-11f being a recurrent nova in a symbiotic system like e.g. RS 
Oph \citep[see also ][]{2003ASPC..303..261H}.
Already
three nova outbursts have been reported from within 2 arcsec of the 
position of M31N~2004-11f (see Table~\ref{tab:rec}). A detailed
astrometry of the three WeCAPP detections shows that the positions of the 
corresponding novae are not the same. However, M31N~2004-11f may still
be a recurrence of nova M31N~1984-07a.
From the WeCAPP data we estimated the decay time from
maximum brightness to a brightness two magnitudes below 
($t_{\rm 2R}$) to 28.4 d.

\section{X-ray observations and methods used for the M~31 nova search}
The search for X-ray counterparts of optical novae in \m31\ reported in PFF2005 
used \xmm\ and \chandra\ observations collected till June 2002. After that 
there were mainly three groups of X-ray observations pointing to the \m31\ 
center area which we extracted from the archive and analyzed for X-ray counterparts of
optical novae.
\begin{itemize}
\item Four 50 ks \chandra\ HRC-I observations to monitor \m31$^*$ (the nuclear
source in M31 corresponding to Cen A$^*$ in the Galaxy) with about one
month spacing which are partly
split in two ObsIDs on the same day (PI Garcia)
\item Several short ($\le 5$ ks) \chandra\ ACIS-I observations of the \m31\ 
bulge area to detect and monitor black hole X-ray transients separated by about
a month (PIs Garcia, Murray)  
\item Four 20 ks \xmm\ observations within four days 
to monitor the low mass X-ray binary 
RX~J0042.6+4115 (PI Barnard) located 1.1\arcmin\ to the west of the \m31\ 
nucleus position
\end{itemize}

We give an observation log of the \chandra\ and \xmm\ observations newly 
analyzed in Tables~\ref{tab:obs_chandra} and \ref{tab:obs_xmm}. There are
several earlier \chandra\ ACIS-I observations to the center of \m31\ from 
August 2002 to May 2004
(ObsID 4360, 4678 -- 4682, 4691). However, no novae were detected 
and we therefore do not show them in 
Table~\ref{tab:obs_chandra}. We checked for offsets of the \chandra\ position
solution from the nominal aspect solution using the point source catalogues of
\citet[][hereafter KGP2002]{2002ApJ...577..738K} and 
\citet[][hereafter K2002]{2002ApJ...578..114K}. No significant
offsets were found ($\le 0.2\arcsec$). For the analysis we merged ObsIDs 6177
and 5926 as well as 6202 and 5927 as they were performed within a day. To be
more sensitive for nova M31N~1995-09b we determined the flux combining all 
HRC observations of Table~\ref{tab:obs_chandra}.
Count rates and 3 $\sigma$ upper limits have been corrected for reduced 
off-axis effective area, assuming a soft
spectrum. At 10\arcmin\ off-axis for the \chandra\ 
high resolution mirror assembly HRMA the area is reduced to about 85\% as 
given in the \chandra\ calibration database v.3.2.1.

For the \xmm\ source detection,
we rejected times with high background (ObsID 0202230301 is totally rejected 
as it shows high background throughout the observation).  To increase the
detection sensitivity we merged the data of ObsIDs 0202230201, 0202230401 and 
0202230501 after correction of the position offset. Within the three day
time-span of the observations one would not expect strong brightness changes of
a nova. For the spectral analysis of nova M31N~1999-10a we used a less stringent
background screening as we were only interested in energies below 1 keV and even
could use part of ObsID 0202230301. Parameters of individual observations are
summarized in Table~\ref{tab:obs_xmm}.
In Tables \ref{tab:novae_old} and 
\ref{tab:novae_new} we assume an average JD for these observations of
2\,453\,205.5.

\begin{table*}
\begin{center}
\caption[]{Log of archival \chandra\ observation to the \m31\
center from July 2004 to February 2005.}
\begin{tabular}{lllrrrr}
\hline\noalign{\smallskip}
\hline\noalign{\smallskip}
\multicolumn{1}{c}{Instr.} &
\multicolumn{1}{c}{ObsID} &\multicolumn{1}{c}{Obs. dates} &
\multicolumn{1}{c}{JD~$^+$} &
\multicolumn{2}{c}{Pointing direction}   &
\multicolumn{1}{c}{$t_{exp}$}\\ 
\noalign{\smallskip}
& & & (2\,450\,000+) &\multicolumn{2}{c}{RA/Dec (J2000)}  
& \multicolumn{1}{c}{(ks)}\\
\noalign{\smallskip}
\multicolumn{1}{c}{(1)} & \multicolumn{1}{c}{(2)} & \multicolumn{1}{c}{(3)} & 
\multicolumn{1}{c}{(4)} & \multicolumn{1}{c}{(5)} & \multicolumn{1}{c}{(6)} & 
\multicolumn{1}{c}{(7)}\\
\noalign{\smallskip}\hline\noalign{\smallskip}
ACIS-I & 4719 & 2004-07-17  & 3204.48 & 0:42:42.98 & 41:16:51.8 &  4.12  \\
ACIS-I & 4720 & 2004-09-02  & 3251.13 & 0:42:43.53 & 41:16:22.6 &  4.11  \\
ACIS-I & 4721 & 2004-10-04  & 3283.40 & 0:42:40.24 & 41:16:14.2 &  4.13  \\
ACIS-I & 4722 & 2004-10-31  & 3309.61 & 0:42:41.10 & 41:15:40.7 &  3.90  \\
ACIS-I & 4723 & 2004-12-05  & 3344.91 & 0:42:49.42 & 41:16:29.6 &  4.04  \\
HRC-I  & 5925 & 2004-12-06  & 3346.6  & 0:42:43.91 & 41:15:54.4 &  46.3  \\
HRC-I  & 6177 & 2004-12-27  & 3366.9  & 0:42:44.22 & 41:15:53.8 &  20.0  \\
HRC-I  & 5926 & 2004-12-27  & 3367.5  & 0:42:44.24 & 41:15:53.3 &  28.3  \\
HRC-I  & 6202 & 2005-01-28  & 3398.7  & 0:42:44.64 & 41:15:53.9 &  18.0  \\
HRC-I  & 5927 & 2005-01-28  & 3399.5  & 0:42:44.65 & 41:15:53.7 &  27.0  \\
HRC-I  & 5928 & 2005-02-21  & 3423.5  & 0:42:44.98 & 41:15:55.0 &  44.9  \\
\noalign{\smallskip}
\hline
\noalign{\smallskip}
\end{tabular}
\label{tab:obs_chandra}
\end{center}
Notes: $^+~~~$: Julian Date at mid of observation\\
\end{table*}

\begin{table*}
\begin{center}
\caption[]{Log of archival \xmm\ observation to the \m31\
center in July 2004.}
\begin{tabular}{llrrrrrrr}
\hline\noalign{\smallskip}
\hline\noalign{\smallskip}
\multicolumn{1}{c}{ObsID} &\multicolumn{1}{c}{Obs. dates} &
\multicolumn{1}{c}{JD~$^+$} &
\multicolumn{2}{c}{Pointing direction} & \multicolumn{1}{c}{Offset~$^*$} 
& \multicolumn{3}{c}{EPIC $t_{exp}^{\dagger}$ (ks)} \\ 
\noalign{\smallskip}
& & (2\,450\,000+) &\multicolumn{2}{c}{RA/Dec (J2000)} & \multicolumn{1}{c}{} 
& \multicolumn{1}{c}{PN}
& \multicolumn{1}{c}{MOS1}
& \multicolumn{1}{c}{MOS2}\\
\noalign{\smallskip}
\multicolumn{1}{c}{(1)} & \multicolumn{1}{c}{(2)} & \multicolumn{1}{c}{(3)} & 
\multicolumn{1}{c}{(4)} & \multicolumn{1}{c}{(5)} & \multicolumn{1}{c}{(6)} & 
\multicolumn{1}{c}{(7)} & \multicolumn{1}{c}{(8)} & \multicolumn{1}{c}{(9)}\\
\noalign{\smallskip}\hline\noalign{\smallskip}
0202230201 & 2004-07-16  & 3203.30 & 0:42:38.58 & 41:16:03.8 & $-1.3,-1.2$ &
16.4 (16.4) & 19.4 & 19.4 \\
0202230301 & 2004-07-17  & 3204.15 & 0:42:38.58 & 41:16:03.8 & $-1.0,-0.9$ &
0.0 (18.6) & 0.0 & 0.0 \\
0202230401 & 2004-07-18  & 3205.62 & 0:42:38.58 & 41:16:03.8 & $-1.7,-1.5$ &
12.4 (12.8) & 17.9 & 17.9 \\
0202230501 & 2004-07-19  & 3206.19 & 0:42:38.58 & 41:16:03.8 & $-1.4,-1.8$ &  
8.0 (19.4) & 10.2 & 10.2 \\
\noalign{\smallskip}
\hline
\noalign{\smallskip}
\end{tabular}
\label{tab:obs_xmm}
\end{center}
Notes:\\
$^+~~~$: Julian Date at mid of observation\\
$^*~~~$: Systematic offset in RA and Dec in arcsec determined from correlations with 2MASS, USNO-B1 and \chandra\ catalogues \\
$^{\dagger}~~~$: Dead time corrected exposure time in units of ks after screening for high
background. In brackets we give the PN exposure time used for the spectral analysis
of nova M31N~1999-10a. All observations in full frame imaging mode with medium filter\\
\end{table*}

\begin{table}
\begin{center}
\caption[]{Count rate conversion factors to un-absorbed fluxes (ECF) into the 0.2--1 keV band 
           for black body models with temperatures of 40~eV and 50~eV 
	   for different instruments and filters, including
	   a Galactic foreground absorption of 6.66\hcm{20}.}
\begin{tabular}{lrrr}
\hline\noalign{\smallskip}
\hline\noalign{\smallskip}
\multicolumn{1}{l}{Detector} & \multicolumn{1}{c}{Filter} &\multicolumn{1}{c}{40 eV} &\multicolumn{1}{c}{50 eV} \\ 
\noalign{\smallskip}
& & \multicolumn{2}{c}{($10^{-11}$ erg cm$^{-2}$ ct$^{-1}$)} \\
\noalign{\smallskip}\hline\noalign{\smallskip}
EPIC PN   & medium & 2.15 & 1.22 \\
EPIC MOS1 & medium & 11.4 & 5.94 \\
EPIC MOS2 & medium & 11.1 & 5.78 \\
Chandra HRC-I  &   & 9.17 & 6.76 \\	  
Chandra ACIS-I &   & 152. & 55.6 \\
\noalign{\smallskip}
\hline
\noalign{\smallskip}
\end{tabular}
\label{tab:ecf}
\end{center}
\end{table}

We calculated intrinsic luminosities or 3$\sigma$ upper limits in the 
0.2--1.0~keV band starting from the 0.2--1~keV count rates or upper limits in
EPIC and ACIS I and the full count rates or upper limits in the \chandra\ HRC
and assuming a black body spectrum and Galactic foreground absorption.
Table~\ref{tab:ecf} gives energy conversion factors for the different
instruments for 40 eV and 50 eV black body temperatures. As one can see
the ECFs strongly change with the softness of the spectrum. Additional absorption within 
\m31\ would heavily change the observed count rate. An extrapolation to the bolometric 
luminosity of a nova at a time is very uncertain and this is even more so as the 
temperatures of novae may vary with time after outburst and from nova to nova and may
well correspond to a spectrum even softer than 30 eV. 

\begin{table}
\begin{center}
\caption[]{Count rate conversion factors from \chandra\ HRC-I to the full energy band of 
           \chandra\ ACIS-I for different spectra derived for \chandra\ Cycle 6 using
	   PIMMS v3.7. For all spectra we assume a Galactic foreground
	   absorption of 6.66\hcm{20}. Model spectra include:  power law 
	   with photon index of 1.7 (PL), thermal bremsstrahlung with a
	   temperature of 1 keV (BR) and three black body spectra (BB) of
	   different temperature.}
\begin{tabular}{rrrrrr}
\hline\noalign{\smallskip}
\hline\noalign{\smallskip}
Spectrum&\multicolumn{1}{c}{PL ($\alpha$)} & \multicolumn{1}{c}{BR (k$T$)} 
&\multicolumn{3}{c}{BB (k$T$)} \\
\noalign{\smallskip}
&1.7 & 1.0 keV&70 eV & 50 eV & 30 eV\\
\noalign{\smallskip}\hline\noalign{\smallskip}
ACIS-I/HRC-I & 1.81 &1.33 &0.28 &0.133 &0.023 \\
\noalign{\smallskip}
\hline
\noalign{\smallskip}
\end{tabular}
\label{tab:HRC_ACIS}
\end{center}
\end{table}

To classify X-ray sources as SSS, we here use \xmm\ EPIC and \chandra\ ASIS-I spectra. Unfortunately, 
many of the novae in 2004 are only detected with the \chandra\ HRC-I and can not be 
classified by this instrument. Nevertheless, several of these novae can 
indirectly be classified
as SSS if they are detected in the \chandra\ HRC-I observation 5925. We then can
make use of the \chandra\ ACIS-I observation 4723 performed 1.7 days earlier. 
Under the assumption that the novae did not change X-ray brightness and spectrum between
these observations, we can check if the corresponding X-ray source had a soft
spectrum by comparing ACIS-I count rates or upper limits with HRC-I count rates. 
Table~\ref{tab:HRC_ACIS} gives count
rate conversion factors from HRC-I to ACIS-I for different spectral
models assuming Galactic foreground absorption. While hard or moderately hard
spectra lead to conversion factors above one, typical supersoft spectra as found
in novae show conversion factors below 0.5. Classification of spectra via this
method are indicated in Table~\ref{tab:novae_old} and \ref{tab:novae_new} with
``(ACIS-I)" under comments.

\section{M~31 optical nova detected with Chandra and/or XMM-Newton}
   \begin{figure*}
   \sidecaption
   \includegraphics[width=12cm,clip]{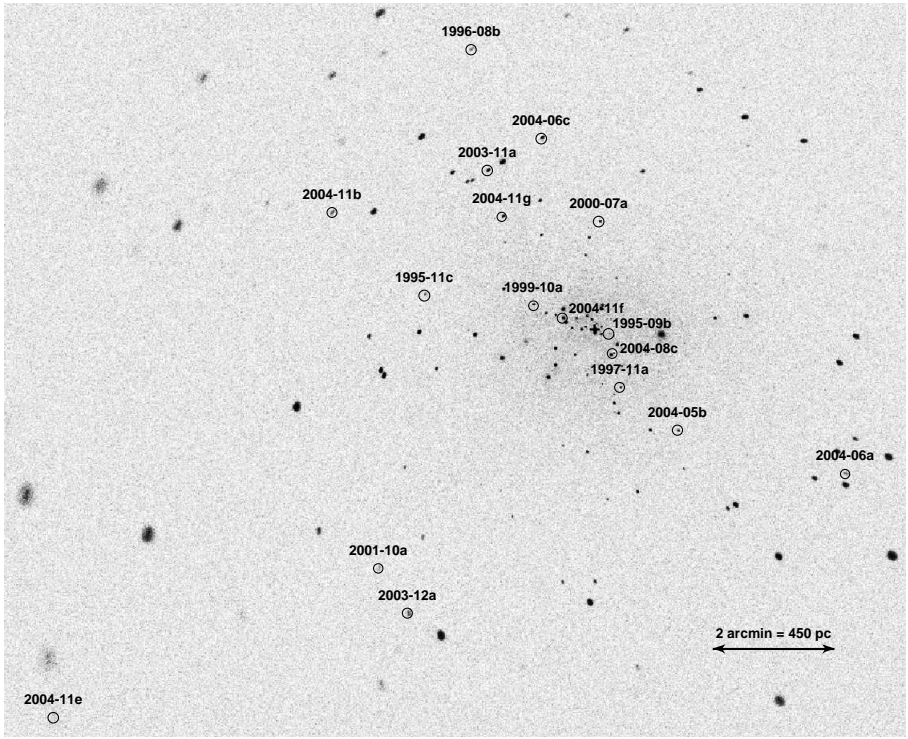}
      \caption{Part of the merged \chandra\ HRC-I image of observations 5925, 6177, 5926,
      6202, 5927, and 5928. 
      Circles with 5" radius indicate positions of optical novae detected in
      these HRC-I observations (see Tables~\ref{tab:novae_old} and 
      \ref{tab:novae_new}). Nova names are given in CBAT nomenclature omitting
      the M31N prefix. The cross between the novae M31N~1995-09b and 
      M31N~2004-11f indicates
      the \m31\ center, the aim point of the observations.
              }
         \label{fig:hrc}
   \end{figure*}
We searched for X-ray emission from nova counterparts 
using two methods with results 
presented in Tables~\ref{tab:novae_old} and \ref{tab:novae_new}, respectively.
The novae in the tables are sorted with ascending time of outburst.

First, we searched for X-ray flux or upper limits
for optical nova counterparts that had been detected in X-rays by PFF2005 and
were still X-ray active at the last observations 
analyzed by PFF2005 (beginning of 2002). We added optical nova counterparts 
that were newly detected in the 2004/5 X-ray observations if the optical
outburst of the nova was before November 2003 
(see Table~\ref{tab:novae_old}). 
In the covered field, the density of novae is rather high and the positions
of novae with outbursts before 1995 are often not as well determined (see
Table A.1). In addition, one would only expect the start of the SSS state later
than 10 years after outburst for novae under extreme conditions 
\citep[low WD mass
and core material with close to solar abundance, see ][]{2006ApJS..167...59H}.
We therefore constrained the search for new X-ray counterparts to novae with 
outburst after 1995 to suppress spurious detections.  

Then we systematically searched for X-ray emission from all optical novae in 
the region covered by the FOV of the X-ray instruments with outburst times between November
2003 and the end of the X-ray coverage in February 2005, giving fluxes or upper 
limits for counterparts (see Table~\ref{tab:novae_new}). To
visualize the large number of novae in the central area of \m31\ that were X-ray active 
during December 2004 to February 2005, we have marked them on a merged \chandra\ HRC-I 
image (Fig. \ref{fig:hrc}). 

\subsection{X-ray detections and upper limits for nova candidates more than a year 
    after outburst}

\begin{table*}
\caption[]{\xmm, \chandra\ and ROSAT measurements of \m31\ optical nova
candidates. 
	   }
\scriptsize
\begin{tabular}{lrrrlrrrrl}
\hline\noalign{\smallskip}
\hline\noalign{\smallskip}
\multicolumn{4}{l}{\normalsize{Optical nova candidate}} 
& \multicolumn{4}{l}{\normalsize{X-ray measurements}} \\
\noalign{\smallskip}\hline\noalign{\smallskip}
\multicolumn{1}{l}{Name} & \multicolumn{1}{c}{RA~~~(h:m:s)$^a$} 
& \multicolumn{1}{c}{Offset$^b$} 
& \multicolumn{1}{c}{JD$^c$} & \multicolumn{1}{l}{Source name$^d$}
& \multicolumn{1}{c}{$D$} 
& \multicolumn{1}{c}{Observation$^e$} 
& \multicolumn{1}{c}{$\Delta T^f$} & \multicolumn{1}{c}{$L_{\rm X}^g$}
& \multicolumn{1}{l}{Comment} \\
M31N & \multicolumn{1}{c}{Dec~(d:m:s)$^a$} &\multicolumn{1}{c}{(\arcmin)}
& \multicolumn{1}{l}{2\,440\,000+} & &(\arcsec)  
& \multicolumn{1}{c}{ID} &\multicolumn{1}{c}{(d)} 
&\multicolumn{1}{c}{(10$^{36}$ erg s$^{-1}$)} & \\ 
\noalign{\smallskip}\hline\noalign{\smallskip}
1994-09a &0:42:42.08&3.87& 9622.5&  	   &   &c3 (EPIC)&2467&$<0.6$&\\
=        &41:12:18.0&&       &J004242.1+411218&0.7&1912 (HRC-I)&2591&$1.2\pm0.4$&K2002\\
AGPV~1576&          &&       &[PFH2005]~313&3.1&c4 (EPIC)&2658&$1.6\pm0.2$&SSS\\
         &          &&       &  	   &   &mrg (EPIC)&3583&$<0.4$\\
\noalign{\medskip}
1995-09b &0:42:43.10&0.24& 9963.5&  	   &   & 268 (HRC-I) &1572&$<7.0$\\
=        &41:16:04.1&&       &  	   &   & 309 (ACIS~S)&1734&$10.2\pm2.7$&\\
RJC99	 &          &&       &  	   &   & 310 (ACIS~S)&1765&$16.1\pm3.6$&\\
Sep-95	 &          &&       &  	   &   &1854 (ACIS~S)&1960&$11.5\pm3.1$&\\
         &          &&       &r1-35	   &1.1&1575 (ACIS~S)&2224&$13.4\pm1.2$&SSS-HR, DKG2004\\
	 &          &&       &J004243.1+411604&0.2&1912 (HRC-I)&2250&$8.8\pm0.8$&K2002\\
	 &          &&       &  	   &   &2905 (HRC-I)&2327&$22.5^{+12.0}_{-8.5}$&\\
	 &          &&       &  	   &   &mrg (HRC-I)&3383&$0.6\pm0.2$ \\
\noalign{\medskip}
1995-11c &0:42:59.3 &2.87&10049.5&  	   &   &400780h (HRI)&  37&$<5.8$&\\
=        &41:16:42  &&       &  	   &   & 268 (HRC-I)&1486&$7.7\pm2.9$&\\
$\rm{[SI2001]}$ &   &&	     &  	   &   & 309 (ACIS~S)&1648&$8.9\pm2.6$&\\
1995-05	 &          &&       &[PFH2005]~369&2.1&c1 (EPIC)&1671&$2.3\pm0.5$&SSS\\
	 &          &&       &  	   &   & 310 (ACIS~S)&1679&$13.8\pm3.3$&\\
	 &          &&       &[PFH2005]~369&2.1&c2 (EPIC)&1857&$3.0\pm0.8$&SSS\\
	 &          &&       &  	   &   &1854 (ACIS~S)&1874&$7.1\pm2.4$&\\
         &          &&       &[PFH2005]~369&2.1&c3 (EPIC)&2040&$4.7\pm0.5$&SSS\\
         &          &&       &r2-63	   &0.8&1575 (ACIS~S)&2138&$11.4\pm1.0$&SSS-HR, DKG2004\\
         &          &&    &J004259.3+411643&1.3&1912 (HRC-I)&2164&$10.5\pm0.9$&K2002\\
         &          &&       &[PFH2005]~369&2.1&c4 (EPIC)&2231&$7.6\pm0.4$&SSS\\
	 &          &&       &  	   &   &2906 (HRC-I)&2378&$16.8^{+9.7}_{-7.1}$&\\	  
         &          &&       &  	   &   &mrg (EPIC)&3156& $3.0\pm0.4$&SSS\\
	 &          &&       &  	   &   &5925 (HRC-I)&3296&$4.6\pm0.9$& \\
	 &          &&       &  	   &   &5926m (HRC-I)&3317&$4.1\pm0.9$& \\
	 &          &&       &  	   &   &5927m (HRC-I)&3349&$3.9\pm0.9$& \\
	 &          &&       &  	   &   &5928 (HRC-I)&3373&$2.6\pm0.8$& \\
\noalign{\medskip}
1996-08b &0:42:55.2 &5.06&10307.5&  	   &   &c3 (EPIC)&1782&$<1.0$&recurrent nova \\
=        & 41:20:46 &&       &  	   &   &1575 (ACIS~S)&1880&$2.2\pm0.8$&\\
GCVS-M31-&          &&    &J004255.3+412045&1.2&1912 (HRC-I)&1906&$2.9\pm0.8$&K2002\\
V0962	 &          &&       &[PFH2005]~359&1.1&c4 (EPIC)&1973&$3.4\pm0.2$&SSS\\
         &          &&       &  	   &   &mrg (EPIC)&2898& $2.0\pm0.3$&SSS\\
	 &          &&       &  	   &   &5925 (HRC-I)&3038&$2.8\pm0.9$& \\
	 &          &&       &  	   &   &5926m (HRC-I)&3059&$4.0\pm0.9$& \\
	 &          &&       &  	   &   &5927m (HRC-I)&3091&$5.6\pm1.0$& \\
	 &          &&       &  	   &   &5928 (HRC-I)&3115&$3.4\pm0.9$& \\
\noalign{\medskip}
1997-08b &0:42:50.5 &8.42&10691.5&  	   &   &c3 (EPIC)&1398&$<1.2$&\\
=        &41:07:48  &&       &  	   &   &1912 (HRC-I)&1522&$<3.7$&\\
$\rm{[SI2001]}$ &   &&	     &[PFH2005]~347&2.3&c4 (EPIC)&1589&$0.7\pm0.2$&SSS\\
1997-09	 &          &&       &  	   &   &mrg (EPIC)&2514&$<0.3$\\
\noalign{\medskip}
1997-11a &0:42:42.13&1.05&10753.6&	   &   &1575 (ACIS~S)&1434&$<0.7$&recurrent nova\\
         &41:15:10.5&&       &  	   &   &1912 (HRC-I)&1461&$<0.7$&\\
	 &          &&       &J004242.1+411511&0.5&5925 (HRC-I)&2593&$3.4\pm0.7$& \\
	 &          &&       &  	   &   &5926m (HRC-I)&2614&$4.4\pm0.7$& \\
	 &          &&       &  	   &   &5927m (HRC-I)&2646&$4.2\pm0.7$& \\
	 &          &&       &  	   &   &5928 (HRC-I)&2670&$4.1\pm0.7$& \\
\noalign{\medskip}
1998-06a &0:43:28.76&10.04&10970.5&  	   &   &c1 (EPIC)&750&$<1.0$&recurrent nova \\
=        &41:21:42.6&&       &  	   &   &c2 (EPIC)&936&$<1.7$& \\
GCVS-M31-&          &&       &[PFH2005]~456&1.1&c3 (EPIC)&1119&$1.3\pm0.4$&SSS \\
V1067	 &          &&       &  	   &   &1912 (HRC-I)&1243&$<5.0$&\\
         &          &&       &[PFH2005]~456&1.1&c4 (EPIC)&1310&$1.7\pm0.3$&SSS \\
         &          &&       &  	   &   &mrg (EPIC)&2235&$<0.4$\\
\noalign{\medskip}
1999-10a & 0:42:49.7&1.08&11454.7&	   &   &1575 (ACIS~S)&733&$<0.7$&\\
         & 41:16:32 &&       &  	   &   &1912 (HRC-I)&760&$<0.8$&\\
         &          &&       &             &   &mrg (EPIC)&1751&$10.5\pm0.9$&SSS\\
	 &          &&       &J004249.7+411633&1.6&5925 (HRC-I)&1892&$21.2\pm1.6$& \\
	 &          &&       &  	   &   &5926m (HRC-I)&1913&$20.8\pm1.5$& \\
	 &          &&       &  	   &   &5927m (HRC-I)&1944&$19.2\pm1.5$& \\
	 &          &&       &  	   &   &5928 (HRC-I)&1969&$20.5\pm1.5$& \\
\noalign{\medskip}
2000-08a &0:42:47.44&1.17&11719.6$^<$&	      &   & 310 (ACIS~S)&  9&$<2.2$&\\
=        &41:15:07.6&&       &  	      &   &1854 (ACIS~S)&203&$<4.2$&\\
WeCAPP-	 &          &&       &  	      &   &1570 (HRC-I)&352&$16.8^{+9.7}_{-7.1}$&\\
N2000-05 &          &&       &      r2-61     &1.7& 1575 (ACIS~S)&468&$9.6\pm1.0$&SSS-HR, DKG2004\\
         &          &&       &J004247.4+411507&0.2&1912 (HRC-I)&494&$5.2\pm0.6$&K2002\\
	 &          &&       &  	   &   &5925 (HRC-I)&1627&$<1.3$ \\
\noalign{\smallskip}
\hline
\noalign{\smallskip}
\end{tabular}
\label{tab:novae_old}
\normalsize
\end{table*}

\begin{table*}
\addtocounter{table}{-1}
\caption[]{continued. 
	   }
\scriptsize
\begin{tabular}{lrrrlrrrrl}
\hline\noalign{\smallskip}
\hline\noalign{\smallskip}
\multicolumn{4}{l}{\normalsize{Optical nova candidate}} 
& \multicolumn{4}{l}{\normalsize{X-ray measurements}} \\
\noalign{\smallskip}\hline\noalign{\smallskip}
\multicolumn{1}{l}{Name} & \multicolumn{1}{c}{RA~~~(h:m:s)$^a$} 
& \multicolumn{1}{c}{Offset$^b$} 
& \multicolumn{1}{c}{JD$^c$} & \multicolumn{1}{l}{Source name$^d$}
& \multicolumn{1}{c}{$D$} 
& \multicolumn{1}{c}{Observation$^e$} 
& \multicolumn{1}{c}{$\Delta T^f$} & \multicolumn{1}{c}{$L_{\rm X}^g$}
& \multicolumn{1}{l}{Comment} \\
M31N & \multicolumn{1}{c}{Dec~(d:m:s)$^a$} &\multicolumn{1}{c}{(\arcmin)}
& \multicolumn{1}{l}{2\,440\,000+} & &(\arcsec)  
& \multicolumn{1}{c}{ID} &\multicolumn{1}{c}{(d)} 
&\multicolumn{1}{c}{(10$^{36}$ erg s$^{-1}$)} & \\ 
\noalign{\smallskip}\hline\noalign{\smallskip}
2000-07a &0:42:43.97&1.78&11753.0$^*$&	   &   &c1 (EPIC)&-32&$<0.7$&\\
=        &41:17:55.5&&       &  	   &   &c2 (EPIC)&154&$<1.4$&\\
WeCAPP-	 &          &&       &  	   &  &1854 (ACIS~S)&170&$7.1\pm2.4$&\\
N2000-03 &          &&       &[PFH2005]~320&1.3&c3 (EPIC)&337&$9.5\pm0.7$&SSS\\
         &          &&       &r2-60	   &0.8&1575 (ACIS~S)&435&$13.4\pm1.2$&SSS-HR, DKG2004\\
         &          &&       &J004243.9+411755&0.3&1912 (HRC-I)&461&$8.4\pm0.8$&K2002\\
         &          &&       &[PFH2005]~320&1.3&c4 (EPIC)&528&$5.5\pm0.4$&SSS\\
         &          &&       &  	   &   &mrg (EPIC)&1453&$13.5\pm0.7$&SSS\\
	 &          &&       &  	   &   &5925 (HRC-I)&1591&$12.0\pm1.2$& \\
	 &          &&       &  	   &   &5926m (HRC-I)&1612&$12.3\pm1.2$& \\
	 &          &&       &  	   &   &5927m (HRC-I)&1644&$12.2\pm1.2$& \\
	 &          &&       &  	   &   &5928 (HRC-I)&1668&$9.6\pm1.1$& \\
\noalign{\medskip}
2001-08d &0:42:34.61&2.76&12150.6$^*$&  	   &   &c3 (EPIC)&-61&$<0.5$&close to\\
=        &41:18:13.0&&       &  	   &   &1575 (ACIS~S)& 37&$<1.9$&[PFH2005]~287\\
WeCAPP-  &          &&       &  	   &$\sim$1&1912 (HRC-I)& 63&$0.6\pm0.3$&\\
N2001-12 &          &&       &  	   &2.5&c4 (EPIC)&130&$0.7\pm0.1$&SSS\\
         &          &&       &  	   &   &mrg (EPIC)&1055&$<0.4$\\
\noalign{\medskip}
2001-10a &0:43:03.31&5.32&12186.4$^*$&	   &   &mrg (EPIC)&1019&$<1.5$&\\
         &41:12:11.5&&       &  	   &   &5925 (HRC-I)&1158&$3.4\pm1.1$& \\
	 &          &&       &  	   &   &5926m (HRC-I)&1179&$2.2\pm0.9$& \\
	 &          &&       &  	   &   &5927m (HRC-I)&1211&$5.2\pm1.3$& \\
	 &          &&       &J004303.2+411211&0.9&5928 (HRC-I)&1235&$3.7\pm1.0$& \\
\noalign{\medskip}
2001-10f &0:41:54.26&12.86   & 12196.3&    &   &1912 (HRC-I)&17&$<6.7$&not [SI2001] 1992-01$^*$\\
         &41:07:23.9&&       &[PFH2005]~191&0.9&c4 (EPIC)&84&$37.0\pm1.7$&SSS\\
         &          &&	     &[PFH2005]~191&0.9&s1 (EPIC)&90&$13.1\pm0.7$&SSS\\
         &          &&       &  	   &   &mrg (EPIC)& 1009&$<0.8$\\
\noalign{\medskip}
2002-01b &0:42:33.89&2.99&12282.3$^<$&	   &	&  c4 (EPIC)&-2&$<0.5$&\\
=        &41:18:23.9&&       &  	   &   &2905 (HRC-I)&8&$<9.1$&\\
WeCAPP-  &          &&       &  	   &$\sim$0.2&2906 (HRC-I)&146&$27.5^{+12.0}_{-9.5}$& \\
N2002-01 &          &&       &  	   &   &mrg (EPIC)&925&$<0.3$ \\
\noalign{\smallskip}
\hline
\noalign{\smallskip}
\end{tabular}

\label{tab:novae_old2}
Notes: \hspace{0.3cm} $^a$: RA, Dec are given in J2000.0 \\
\hspace*{1.cm} $^b$: projected distance from \m31\ nucleus position
\citep[RA$_\mathrm{J2000} = 00^h42^m44\fs324, \delta_\mathrm{J2000} =
+41\degr16\arcmin08\farcs53$; ][]{1992ApJ...390L...9C}\\
\hspace*{1.cm} $^c$: $^*$ indicates well defined start date of optical outburst, 
$^<$ outburst start before, else badly defined (see text) \\
\hspace*{1.cm} $^d$: full source names from  K2002 are CXOM31~Jhhmmss.s+ddmmss \\
\hspace*{1.cm} $^e$: for \xmm\ c1 corresponds to ObsID 0112570401, c2 to 
0112570601, c3 to 0112570701, c4 to 0112570101 and s1 to 
0112570201 (see PFF2005),\\ 
\hspace*{1.3cm}``mrg (EPIC)" indicates the merged data of ObsID 0202230201,
 0202230401, and 0202230501\\ 
\hspace*{1.3cm}for \chandra\ we give ObsID and camera used (see text). 5926m
 indicate the merged data of ObsIDs 6177 and 5926, 5927m those of 6202 and 5927,\\
\hspace*{1.3cm}``mrg (HRC-I)" indicates the merged data of ObsIDs 5925--5928,
6177 and 6202\\
\hspace*{1.cm} $^f$: time after observed start of optical outburst\\
\hspace*{1.cm} $^g$: un-absorbed luminosity in 0.2--1.0 keV band assuming a 50 eV 
black body spectrum with Galactic absorption, upper limits are 3$\sigma$ \\
\hspace*{1.cm} $^*$: wrong identification by PFF2005 \citep[see ][]{2006IBVS.5720....1S} \\
\normalsize
\end{table*}

In the 2004/2005 data, we searched for X-ray emission from the eleven 
\m31\ optical novae that did not finish 
their SSS state before the last X-ray observations reported by PFF2005, i.e. 
by January 2002. In addition we
detected three novae with optical outbursts from 1997 to 2001 that got X-ray
active after January 2002. 
Results on these novae are summarized in 
Table~\ref{tab:novae_old}. We give nova name (with reference to the name used by
PFF2005) in column 1, optical position (J2000.0, col. 2), distance from \m31\
center (3), and Julian date JD of start of outburst
 (4). We indicate if the time of outburst is well defined (to better
than 5 days), or if the outburst occurred most likely before the given epoch 
or is not well defined.  As X-ray information we give the
name of the source (5), distance $D$ between X-ray and optical position (6),
observation number (7), days since
optical nova outburst (8), X-ray luminosity in the 0.2--1.0 keV band as
described above (9), and comments like reference for detection, nova type, and 
SSS classification (10). For novae close to the \m31\ center we only give
\chandra\ fluxes and upper limits as \xmm\ in most cases can not resolve the 
sources from bright nearby sources and diffuse emission.

From the eleven novae from the PFF2005 list in Table~\ref{tab:novae_old}, seven 
were no longer X-ray active in 2004. Due to the long gap in X-ray 
observations to the \m31\ center that were sensitive to detect supersoft X-ray
emission, the time of the end of the X-ray activity of these novae is not well 
constrained. For three of the optical novae in the table we give slightly 
improved optical positions and/or time of start of optical outburst compared to
PFF2005 (M31N~2000-08a, M31N~2000-08d, M31N~2002-01b). For the SSS [PFH2005]~191
we now give the parameters for the new nova identification of 
\citet{2006IBVS.5720....1S} showing that the nova was already detected as a 
bright SSS 84 days after the optical outburst. 

Four of the novae from the PFF2005 list (Nova M31N~1995-09b, 
M31N~1995-11c, M31N~1996-08b, M31N~2000-07a) 
are still detected in 2004. Nova M31N~1995-09b has dropped
in brightness by more than a factor of 10 since the last detection in January 
2002. Merging all 2004/5 HRC-I observations, 
it is detected with a significance of $\sim 3 \sigma$  9.3 years after the 
optical outburst. The X-ray brightness of nova M31N~1995-11c has dropped by 
about a factor of three since the last detection in June 2002, however, is
clearly visible in all \xmm\ EPIC and \chandra\ HRC-I observations 2004/5, now
9.2 years after the optical outburst. Nova M31N~1996-08b, now 8.5 years
after the optical outburst, stayed at the
same X-ray brightness in 2004/5 or even slightly increased in brightness.
The start of the SSS phase for nova M31N~2000-07a was constrained
to 154--170 days after the optical nova outburst. It is detected at similar
X-ray brightness in 2004/5, more than 4.5 yr after the optical outburst.

\begin{figure}
   \resizebox{\hsize}{!}{\includegraphics[angle=90,clip]{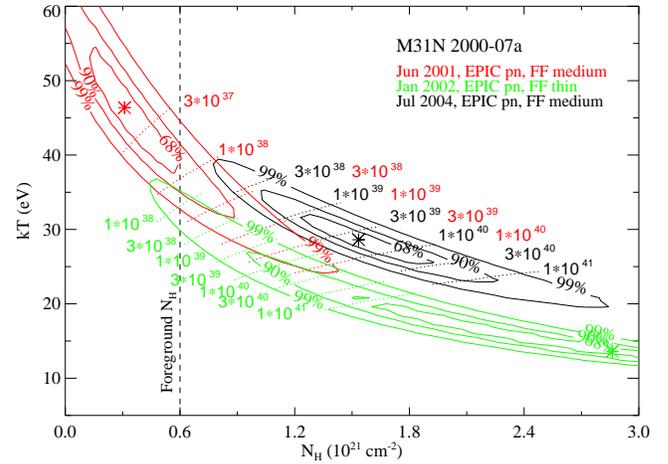}}
     \caption[]{ Column density -- temperature confidence contours inferred 
               from the absorbed black body model fit to the \xmm\ EPIC PN 
               spectra of M31N~2000-07a. 
	       The formal best 
	       fit parameters are indicated by stars. Also drawn are lines of 
	       constant bolometric luminosity (in erg s$^{-1}$) 
	       for a distance of 780 kpc. The 
	       vertical dashed line indicates the Galactic foreground absorption 
	       in the direction of M31. Contours for the observations in June
	       2001, January 2002 and July 2004 are coded in dark and light
	       grey and black [see the electronic edition of the Journal for a
	       color version of this figure].
     }
    \label{pncont_2000_07a} 
\end{figure}
We investigated for which of these novae we could compare the X-ray spectrum
in 2004 to earlier spectra. For M31N~1995-09b, M31N~1995-11c 
and  M31N~1996-08b, the number of counts above background is too low 
for detailed spectral fitting. However, for  M31N~2000-07a,
we collected in total about 900 source counts with
the EPIC PN detector from ObsIDs 0202230201, 0202230301,
0202230401, 0202230501 which allowed spectral fitting and comparison to the
spectral parameters derived for June 2001 and January 2002. In contrast to
PFF2005, we use for this comparison just the EPIC PN detector due to its
stable low energy response with time. We only rejected times of
strong background flares in the (0.2 -- 1.0) keV band to derive spectra for
the July 2004 observations. In this way, 
significantly more time of the observations was usable than for the source 
detection procedures
for which images with low background in all bands were needed 
(see Table~\ref{tab:obs_xmm}). The spectra of the July 2004 observations were 
binned to contain at least 20
counts per bin and simultaneously fitted by an absorbed 
\citep[tbabs in XSPEC,][]{2000ApJ...542..914W} blackbody model. June 2001 and 
January 2002 data were analyzed in the same way. Confidence
contours for absorption column density and blackbody temperature are shown in
Fig~\ref{pncont_2000_07a}. If we assume that the supersoft emission originates
from the surface of a WD in \m31, fit results with \nh\ below the Galactic 
foreground and also results leading to bolometric luminosities above the 
Eddington luminosity of a WD can be excluded \citep[at maximum 
3.5\ergs{38}  for a WD with
a mass at the Chandrasekhar limit, i.e. 1.4 \msun, and He-rich atmosphere, see
e.g. ][]{1993SSRv...62..223L}.

The three novae counterparts  that were not yet in the SSS state during the 
observations analyzed by PFF2005
(M31N~1997-11a, M31N~1999-10a, M31N~2001-10a) 
will be discussed in more detail in the following sub-sections.

\subsubsection{Nova M31N~1997-11a}\label{1997-11a}
This nova candidate was reported by \citet[][hereafter RJC99]{1999AAS...195.3608R} from
one H$\alpha$ image on November 18, 1997. The object is also reported by 
\citet[][hereafter SI2001]{2001ApJ...563..749S} as nova 1997-07 from
one H$\alpha$ image on November 2, 1997. The nova was detected in the WeCAPP
program as N1997-03 in the R band on November 1, 1997 and in the I band one day
later.
According to SI2001 the nova coincides in position with nova
M31N~1982-08b and
is therefore classified as recurrent nova ($\Delta T \sim 15$ yr). 
In X-rays, a counterpart
was first detected in the \chandra\ HRC-I observation 5925 about 7.1
years after the last reported optical outburst. The source is detected
throughout HRC-I observation 5928. As it is rather faint the upper limit from
the \chandra\ ACIS-I observation 4723 only leads to an ACIS-I/HRC-I 
count rate factor less than 1.3. This does not significantly constrain the X-ray spectrum. We
therefore can not decide if it is a SSS.

\subsubsection{Nova M31N~1999-10a}\label{1999-10a}
\begin{figure}
   \resizebox{\hsize}{!}{\includegraphics[angle=-90,clip]{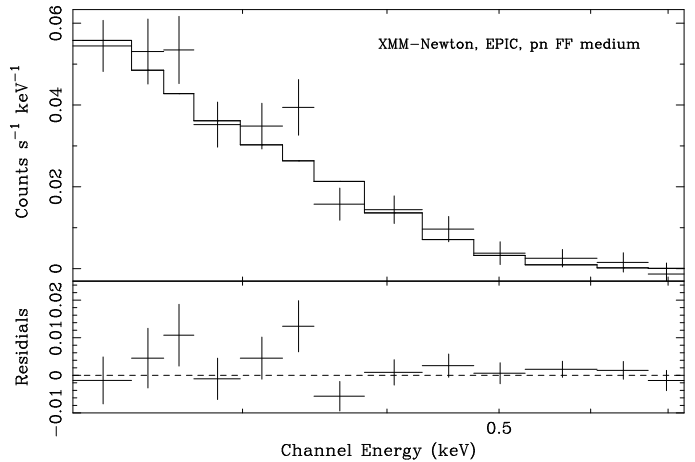}}
     \caption[]{Combined \xmm\ EPIC PN spectrum of nova M31N~1999-10a. 
     The absorbed black body 
     fit to the data (see Sect.~\ref{1999-10a}) is shown in the upper panel. 
     }
    \label{pnspec_1999_10a} 
\end{figure}
\begin{figure}
   \resizebox{\hsize}{!}{\includegraphics[angle=90,clip]{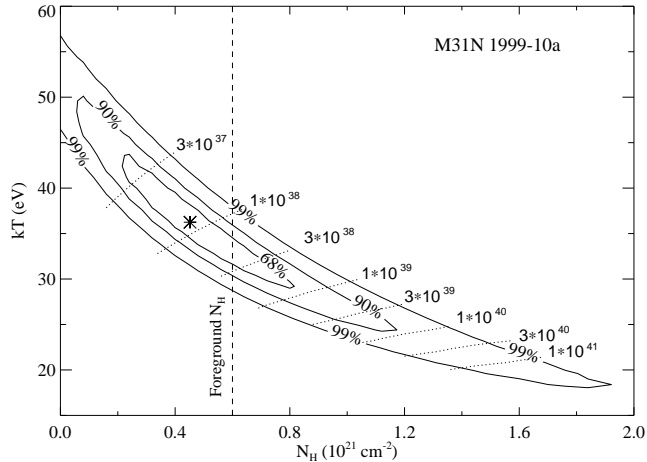}}
     \caption[]{ Column density -- temperature confidence contours inferred 
               from the fit to the \xmm\ EPIC PN 
               spectra of M31N~1999-10a  
	       (see Fig.~\ref{pnspec_1999_10a}). The formal best 
	       fit parameters are indicated by stars. Also drawn are lines of 
	       constant bolometric luminosity and for the Galactic foreground 
	       absorption 
	       (see Fig.~\ref{pncont_2000_07a}).
     }
    \label{pncont_1999_10a} 
\end{figure}

This nova candidate was reported by \citet[][hereafter FCL99]{1999IAUC.7272....3F}. 
FCL99 constrain the date of outburst to within a day (October 2, 1999) and 
confirmed the nova identification with an optical spectrum.
An X-ray counterpart was detected in the
\xmm\ observations in July 2004, 4.8 years after the optical outburst.
From the X-ray hardness ratios, the source  can be classified 
as SSS. It stayed X-ray active till end of February 2005.

An X-ray spectrum was obtained from \xmm\ EPIC PN ObsIDs 0202230201, 0202230301,
0202230401, 0202230501 (giving in total 
$\sim540$ counts from the source). 
The spectra of the four observations were again (see description for
M31N~2000-07a)
simultaneously fitted by an absorbed blackbody model giving 
a best fit of
\nh$\,=\left(4^{+8}_{-4}\right)\times10^{20}\,$\cm-2 and $kT =
\left(36\pm13\right)$~eV. For plotting, the four spectra were summed to one
spectrum (see Fig.~\ref{pnspec_1999_10a}). To sum them up, we used the binning 
of the spectrum of 0202230201. Confidence
contours for absorption column density and blackbody temperature are shown in
Fig~\ref{pncont_1999_10a}.

\subsubsection{Nova M31N~2001-10a}
This nova candidate was first reported by \citet{2001IAUC.7729....2L} and 
spectroscopically confirmed as nova in
\m31\ by \citet{2001IAUC.7738....3F} with strong Balmer and Fe II emission 
lines. Light curves identifying the nova as
moderately fast, are reported from the POINT-AGAPE microlensing survey 
\citep{2004MNRAS.351.1071A,2004MNRAS.353..571D} and from the Nainital
microlensing survey  \citep{2004A&A...415..471J}. It is catalogued as N2001-13
by WeCAPP with a well defined position and date of outburst. 

In X-rays, a counterpart 
was first detected in the \chandra\ HRC-I observation 5925 about 3.2
years after the optical outburst. The source is detected
throughout HRC-I observation 5928. As it is rather faint, the upper limit from
the \chandra\ ACIS-I observation 4723 only leads to an ACIS-I/HRC-I 
count rate factor less than 2.4. This does not constrain the X-ray spectrum. We
therefore can not decide if the source has a supersoft spectrum.

\subsection{Nova candidates detected in 2004/2005 within about a year after
outburst}\label{sect:novae_new}
   \begin{figure*}
  \centering
   \includegraphics[width=17.5cm]{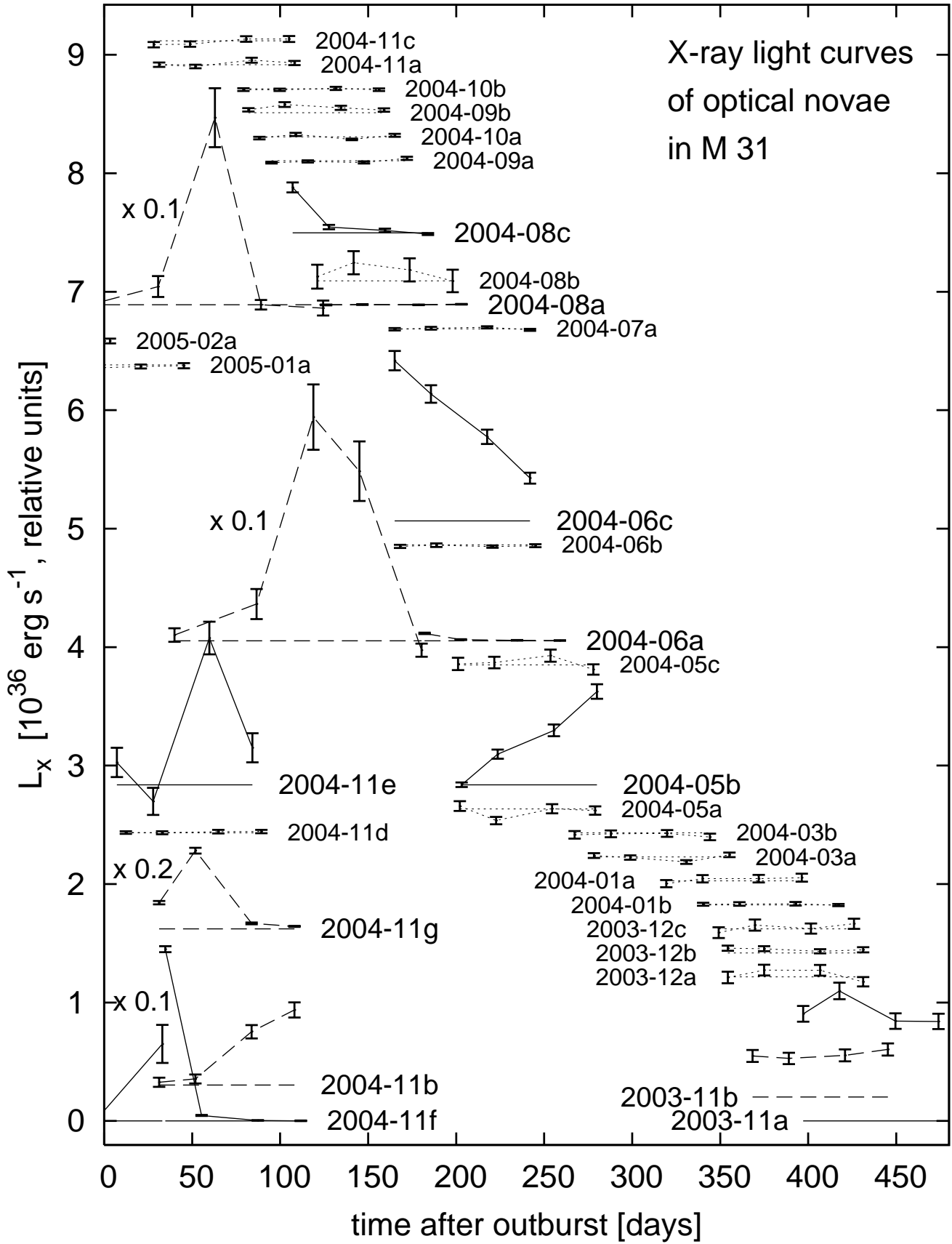}
      \caption{X-ray light curves of optical nova
      counterparts observed by the \chandra\ HRC-I and ACIS-I instruments
      during 2004/5 within about a year after outburst. 
      Detected sources are indicated by solid or dashed 
      lines, not detected novae by dotted lines. 
      Zero level for the individual 
      light curves are indicated and have been shifted for clarity. They are
      labeled with the nova names (following CBAT
      nomenclature) using a bigger font for detected novae
        (see text and Table~\ref{tab:novae_new}).       }
         \label{fig:LCs}
   \end{figure*}
We searched for X-ray emission from nova candidates detected in 2004/2005 within about 
a year after outburst. We determined count rates and upper limits for the HRC-I
observations and \xmm\ observations. These results are given in 
Table~\ref{tab:novae_new} using the same layout as for 
Table~\ref{tab:novae_old} (see above). We also give 
count rates and upper limits for the ACIS-I observations after nova outburst
if a counterpart is detected in at least one ACIS-I observation. 
34 novae have been
reported in the covered field with optical outburst between November 2003 and  
before the last X-ray observation. 

Figure~\ref{fig:LCs} shows the corresponding \chandra\ HRC-I (and ACIS-I) 
light curves plotting un-absorbed X-ray luminosities in the
0.2--1 keV band (assuming a 50 eV black body spectrum with Galactic absorption)
against time after optical nova outburst. We included light curves of detected
nova counterparts and non-detections plotting for each nova data points with 
1 $\sigma$ errors connected by a line. Zero
levels for the individual light curves are also given. The light curves 
have been shifted to avoid overlaps. For better visibility  light curves of
detected candidates are indicated by solid or dashed lines, not detected novae
by dotted lines. 

Two nova candidates are excluded from the list. The first is
M31N~2004-12a, which exploded close to
the bright persistent source [PFH2005] 269 
corresponding to CXOM31 J004238.2+411000 (K2002) or r3-36 (KGP2002) at an offset
of 6\farcm9 from the \m31\ center. Due to the far off-axis distance X-ray
emission from the nova - if any - could not be separated from emission from the 
bright source. 
The second is M31N~2004-02a.
As discussed in Section \ref{sect:opt}, M31N~2004-02a and 
M31N~2004-11f are only separated by $\sim$2\arcsec. We detect X-ray
emission from within 0\farcs3 from the position of M31N~2004-11f and
therefore connect the emission to this nova candidate and not to M31N~2004-02a
(see Table~\ref{tab:novae_new}). 

From the remaining 32 nova candidates, eleven have counterparts in X-rays 
detected as transient sources till end
of February 2005. These nova candidates will be discussed in more detail in the following 
subsections. This leaves 21 nova candidates that have not been detected in X-rays.
Possible reasons for the lack of detection will be discussed later.
 
\begin{table*}
\caption[]{X-ray detections and upper limits of \m31\ 
optical nova candidates within about a year after outburst using \chandra\ and \xmm\ 
observations from July 2004 to February 2005. 
	   }
\scriptsize
\begin{tabular}{lrrrlrrrrl}
\hline\noalign{\smallskip}
\hline\noalign{\smallskip}
\multicolumn{4}{l}{\normalsize{Optical nova candidate}} 
& \multicolumn{4}{l}{\normalsize{X-ray measurements}} \\
\noalign{\smallskip}\hline\noalign{\smallskip}
\multicolumn{1}{l}{Name} & \multicolumn{1}{c}{RA~~~(h:m:s)$^a$} 
& \multicolumn{1}{c}{Offset$^b$} 
& \multicolumn{1}{c}{JD$^c$} & \multicolumn{1}{l}{Source name$^d$}
& \multicolumn{1}{c}{$D$} 
& \multicolumn{1}{c}{Observation$^e$} 
& \multicolumn{1}{c}{$\Delta T^f$} & \multicolumn{1}{c}{$L_{\rm X}^g$}
& \multicolumn{1}{l}{Comment} \\
M31N & \multicolumn{1}{c}{Dec~(d:m:s)$^a$} &\multicolumn{1}{c}{(\arcmin)}
& \multicolumn{1}{l}{2\,450\,000+} & &(\arcsec)  
& \multicolumn{1}{c}{ID} &\multicolumn{1}{c}{(d)} 
&\multicolumn{1}{c}{(10$^{36}$ erg s$^{-1}$)} & \\ 
\noalign{\smallskip}\hline\noalign{\smallskip}
2003-11a &0:42:53.78&3.17&2948.5$^*$&&&mrg (EPIC)&257&$<2.5$& \\
         &41:18:46.2&&       &J004253.7+411846&0.2&5925 (HRC-I)&397&$22.7\pm1.7$& SSS (ACIS-I)\\
	 &          &&	        &  	   &   &5926m (HRC-I)&418&$27.6\pm1.8$& \\
	 &          &&       &  	   &   &5927m (HRC-I)&450&$21.2\pm1.6$& \\
	 &          &&       &  	   &   &5928 (HRC-I)&474&$21.1\pm1.6$& \\
\noalign{\medskip}
2003-11b &0:43:00.76&5.62&2973.4$^*$ &&&mrg (EPIC)&232&$<0.9$& \\
         &41:11:26.9&&       & J004300.7+411126&0.5&5925 (HRC-I)&372&$9.0\pm1.3$& soft (ACIS-I)\\  
	 &          &&       &  	   &   &5926m (HRC-I)&393&$8.4\pm1.2$& \\
	 &          &&       &  	   &   &5927m (HRC-I)&425&$9.1\pm1.3$& \\
	 &          &&       &  	   &   &5928 (HRC-I)&449&$10.4\pm1.3$& \\
\noalign{\medskip}
2003-12a &0:43:04.73&5.38&2992.3 &  	   &   &mrg (EPIC)&213&$<0.6$& \\
         &41:12:21.9&&       &  	   &   &5925 (HRC-I)&353&$<3.7$& \\
	 &          &&       &  	   &   &5926m (HRC-I)&374&$<5.0$& \\
	 &          &&       &  	   &   &5927m(HRC-I)&406&$<5.0$& \\
	 &          &&       &  	   &   &5928 (HRC-I)&430&$<3.1$& \\
\noalign{\medskip}
2003-12b &0:42:54.14&2.07&2992.3 &  	   &   &mrg (EPIC)&213&$<0.3$& \\
         &41:15:12.2&&       &  	   &   &5925 (HRC-I)&352&$<2.6$& \\
	 &          &&       &  	   &   &5926m (HRC-I)&373&$<2.4$& \\
	 &          &&       &  	   &   &5927m (HRC-I)&405&$<1.6$& \\
	 &          &&       &  	   &   &5928 (HRC-I)&429&$<2.2$& \\
\noalign{\medskip}
2003-12c &0:42:53.24&6.67&2994.2 &  	   &   &mrg (EPIC)&211&$<1.7$& \\
         &41:22:35.9&&       &  	   &   &5925 (HRC-I)&351&$<3.7$& \\
	 &          &&       &  	   &   &5926m (HRC-I)&372&$<4.6$& \\
	 &          &&       &  	   &   &5927m (HRC-I)&404&$<3.6$& \\
	 &          &&       &  	   &   &5928 (HRC-I)&428&$<4.6$& \\
\noalign{\medskip}
2004-01b &0:42:41.19&0.71&3006.2 &  	   &   &5925 (HRC-I)&339&$<1.0$& \\
         &41:15:45.0&&       &  	   &   &5926m (HRC-I)&360&$<1.3$& \\
	 &          &&       &  	   &   &5927m (HRC-I)&392&$<1.3$& \\
	 &          &&       &  	   &   &5928 (HRC-I)&416&$<0.7$& \\
\noalign{\medskip}
2004-01a &0:43:08.65&4.60&3027.2 &  	   &   &mrg (EPIC)&178&$<0.8$& \\
         &41:15:35.4&&       &  	   &   &5925 (HRC-I)&318&$<2.4$& \\
	 &          &&       &  	   &   &5926m (HRC-I)&339&$<2.9$& \\
	 &          &&       &  	   &   &5927m (HRC-I)&371&$<2.9$& \\
	 &          &&       &  	   &   &5928 (HRC-I)&395&$<3.2$& \\
\noalign{\medskip}
2004-03a &0:42:36.21&1.61&3068.3$^<$ &  	  &   &mrg (EPIC)&137&$<2.7$& \\
         &41:15:37.9&&       &  	   &   &5925 (HRC-I)&278&$<1.8$& \\
	 &          &&       &  	   &   &5926m (HRC-I)&299&$<1.4$& \\
	 &          &&       &  	   &   &5927m (HRC-I)&331&$<1.2$& \\
	 &          &&       &  	   &   &5928 (HRC-I)&355&$<1.8$& \\
\noalign{\medskip}
2004-03b &0:43:06.72&5.92&3079.3 &  	  &   &mrg (EPIC)&126&$<0.7$& \\
         &41:11:58.5&&       &  	   &   &5925 (HRC-I)&267&$<2.5$& \\
	 &          &&       &  	   &   &5926m (HRC-I)&288&$<2.2$& \\
	 &          &&       &  	   &   &5927m (HRC-I)&320&$<2.3$& \\
	 &          &&       &  	   &   &5928 (HRC-I)&344&$<2.1$& \\
\noalign{\medskip}
2004-05b &0:42:37.04&2.16&3143.6 &  	   &   &mrg (EPIC)& 62&$<0.9$& \\
         &41:14:28.5&&       &  	   &   &5925 (HRC-I)&202&$<1.4$& \\
	 &          &&       &  	   &   &5926m (HRC-I)&223&$6.4\pm0.9$& \\
	 &          &&       &  	   &   &5927m (HRC-I)&255&$11.3\pm1.2$& \\
	 &          &&       &J004237.0+411428&0.6&5928 (HRC-I)&279&$19.4\pm1.5$& \\
\noalign{\medskip}
2004-05a &0:42:37.55&6.01&3144.6 &  	  &   &mrg (EPIC)& 61&$<1.0$& \\
         &41:10:16.4&&       &  	   &   &5925 (HRC-I)&201&$<3.8$& \\
	 &          &&       &  	   &   &5926m (HRC-I)&222&$<2.5$& \\
	 &          &&       &  	   &   &5927m (HRC-I)&254&$<3.1$& \\
	 &          &&       &  	   &   &5928 (HRC-I)&278&$<2.8$& \\
\noalign{\medskip}
2004-05c &0:43:04.04&8.43&3145.6 &  	  &   &mrg (EPIC)& 61&$<0.6$& \\
         &41:23:42.6&&       &  	   &   &5925 (HRC-I)&200&$<4.6$& \\
	 &          &&       &  	   &   &5926m (HRC-I)&221&$<4.6$& \\
	 &          &&       &  	   &   &5927m (HRC-I)&253&$<6.4$& \\
	 &          &&       &  	   &   &5928 (HRC-I)&277&$<3.6$& \\
\noalign{\medskip}
2004-06a &0:42:22.31&4.78&3164.5 &  	  &   &4719 (ACIS-I)&40&$<7.3$& \\
         &41:13:44.9&&       &  	   &   &mrg (EPIC)& 41&$<0.7$& \\
	 &          &&       &  	   &   &4720 (ACIS-I)&87&$10.2\pm4.2$& \\
	 &          &&       &  	   &   &4721 (ACIS-I)&119&$62.1\pm9.0$& SSS\\
	 &          &&       &  	   &   &4722 (ACIS-I)&145&$47.1\pm8.3$& \\
	 &          &&       &  	   &   &4723 (ACIS-I)&180&$<5.4$& \\
	 &          &&       &J004222.3+411345&1.0&5925 (HRC-I)&181&$16.2\pm1.5$& \\
	 &          &&       &  	   &   &5926m (HRC-I)&202&$2.8\pm0.8$& \\
	 &          &&       &  	   &   &5927m (HRC-I)&234&$<3.1$& \\
	 &          &&       &  	   &   &5928 (HRC-I)&258&$<2.6$& \\
\noalign{\smallskip}
\hline
\noalign{\smallskip}
\end{tabular}
\label{tab:novae_new}
\normalsize
\end{table*}

\begin{table*}
\addtocounter{table}{-1}
\caption[]{continued. 
	   }
\scriptsize
\begin{tabular}{lrrrlrrrrl}
\hline\noalign{\smallskip}
\hline\noalign{\smallskip}
\multicolumn{4}{l}{\normalsize{Optical nova candidate}} 
& \multicolumn{4}{l}{\normalsize{X-ray measurements}} \\
\noalign{\smallskip}\hline\noalign{\smallskip}
\multicolumn{1}{l}{Name} & \multicolumn{1}{c}{RA~~~(h:m:s)$^a$} 
& \multicolumn{1}{c}{Offset$^b$} 
& \multicolumn{1}{c}{JD$^c$} & \multicolumn{1}{l}{Source name$^d$}
& \multicolumn{1}{c}{$D$} 
& \multicolumn{1}{c}{Observation$^e$} 
& \multicolumn{1}{c}{$\Delta T^f$} & \multicolumn{1}{c}{$L_{\rm X}^g$}
& \multicolumn{1}{l}{Comment} \\
M31N & \multicolumn{1}{c}{Dec~(d:m:s)$^a$} &\multicolumn{1}{c}{(\arcmin)}
& \multicolumn{1}{l}{2\,450\,000+} & &(\arcsec)  
& \multicolumn{1}{c}{ID} &\multicolumn{1}{c}{(d)} \\
\noalign{\smallskip}\hline\noalign{\smallskip}
2004-06b &0:42:41.30&2.15&3178.5 &  	  &   &mrg (EPIC)& 27&$<0.2$& \\
         &41:14:04.2&&       &  	   &   &5925 (HRC-I)&168&$<1.0$& \\
	 &          &&       &  	   &   &5926m (HRC-I)&189&$<1.1$& \\
	 &          &&       &  	   &   &5927m (HRC-I)&221&$<0.8$& \\
	 &          &&       &  	   &   &5928 (HRC-I)&245&$<0.9$& \\
\noalign{\medskip}
2004-06c &0:42:49.02&3.28&3181.5 &  	  &   &mrg (EPIC)& 24&$<1.0$& \\
         &41:19:17.8&&       &J004249.0+411918&0.9&5925 (HRC-I)&165&$33.9\pm2.0$& SSS (ACIS-I)\\
	 &          &&       &  	   &   &5926m (HRC-I)&186&$26.9\pm1.9$& \\
	 &          &&       &  	   &   &5927m (HRC-I)&218&$17.8\pm1.5$& \\
	 &          &&       &  	   &   &5928 (HRC-I)&242&$9.1\pm1.2$& \\
\noalign{\medskip}
2004-07a &0:42:43.88&1.44&3181.5 &  	  &   &mrg (EPIC)& 24&$<0.8$& \\
         &41:17:35.0&&       &  	   &   &5925 (HRC-I)&165&$<0.7$& \\
	 &          &&       &  	   &   &5926m (HRC-I)&186&$<0.9$& \\
	 &          &&       &  	   &   &5927m (HRC-I)&218&$<1.0$& \\
	 &          &&       &  	   &   &5928 (HRC-I)&242&$<0.6$& \\
\noalign{\medskip}
2004-08a &0:42:20.62&4.45&3220.5$^<$ &  	  &   &4719 (ACIS-I)&-15&$<4.0$& \\
         &41:16:09.5&&       &  	   &   &4720 (ACIS-I)&32&$<13.8$& \\
	 &          &&       &J004220.7+411608&1.8&4721 (ACIS-I)&64&$51.9\pm8.2$& SSS\\
	 &          &&       &  	   &   &4722 (ACIS-I)&90&$<4.0$& \\
	 &          &&       &  	   &   &4723 (ACIS-I)&125&$<6.2$& \\
	 &          &&       &  	   &   &5925 (HRC-I)&126&$<2.1$& \\
	 &          &&       &  	   &   &5926m (HRC-I)&147&$<2.8$& \\
	 &          &&       &  	   &   &5927m (HRC-I)&179&$<2.0$& \\
	 &          &&       &  	   &   &5928 (HRC-I)&203&$<3.4$& \\
\noalign{\medskip}
2004-08b &0:43:26.84&8.01&3225.5$^*$ &  	  &   &5925 (HRC-I)&121&$<9.2$& \\
         &41:16:40.8&&       &  	   &   &5926m (HRC-I)&142&$<12.2$& \\
	 &          &&       &  	   &   &5927m (HRC-I)&174&$<10.5$& \\
	 &          &&       &  	   &   &5928 (HRC-I)&198&$<7.8$& \\
\noalign{\medskip}
2004-08c &0:42:42.77&0.49&3239.5 &J004242.7+411545&0.3&5925 (HRC-I)&107&$9.3\pm1.0$& SSS (ACIS-I)\\
         &41:15:44.7&&       &  	   &   &5926m (HRC-I)&128&$1.2\pm0.4$& \\
	 &          &&       &  	   &   &5927m (HRC-I)&160&$<1.6$& \\
	 &          &&       &  	   &   &5928 (HRC-I)&184&$<0.7$& \\
\noalign{\medskip}
2004-09a &0:42:40.27&1.62&3251.5$^*$ &  	  &   &5925 (HRC-I)&95&$<0.5$& \\
         &41:14:42.5&&       &  	   &   &5926m (HRC-I)&116&$<0.7$& \\
	 &          &&       &  	   &   &5927m (HRC-I)&148&$<0.7$& \\
	 &          &&       &  	   &   &5928 (HRC-I)&172&$<1.5$& \\
\noalign{\medskip}
2004-10a &0:42:51.84&1.42&3258.4 &  	   &   &5925 (HRC-I)&88&$<0.8$& \\
         &41:16:18.2&&       &  	   &   &5926m (HRC-I)&109&$<1.5$& \\
	 &          &&       &  	   &   &5927m (HRC-I)&141&$<0.5$& \\
	 &          &&       &  	   &   &5928 (HRC-I)&165&$<1.2$& \\
\noalign{\medskip}
2004-09b &0:42:44.45&0.04&3264.6 &  	   &   &5925 (HRC-I)&82&$<1.6$& \\
         &41:16:10.5&&       &  	   &   &5926m (HRC-I)&103&$<3.2$& \\
	 &          &&       &  	   &   &5927m (HRC-I)&135&$<2.3$& \\
	 &          &&       &  	   &   &5928 (HRC-I)&159&$<1.6$& \\
\noalign{\medskip}
2004-10b &0:42:47.24&0.60&3267.4 &  	   &   &5925 (HRC-I)&79&$<0.9$& \\
         &41:15:54.5&&        & 	   &   &5926m (HRC-I)&100&$<0.8$& \\
	 &          &&        & 	   &   &5927m (HRC-I)&132&$<1.0$& \\
	 &          &&        & 	   &   &5928 (HRC-I)&156&$<0.8$& \\
\noalign{\medskip}
2004-11f &0:42:47.15&0.56&3311.8 &  	   &   &4722 (ACIS-I)&-1&$<67.2$&recurrent?\\
         &41:16:19.8&&       &  	   &   &4723 (ACIS-I)&34&$20.3\pm5.0$& \\
	 &          &&       &J004247.1+411620&0.3&5925 (HRC-I)&35&$353.3\pm6.2$& SSS (ACIS-I)\\
	 &          &&       &  	   &   &5926m (HRC-I)&55&$11.4\pm1.2$& \\
	 &          &&       &  	   &   &5927m (HRC-I)&87&$1.2\pm0.6$& \\
	 &          &&       &  	   &   &5928 (HRC-I)&112&$<2.0$& \\
\noalign{\medskip}
2004-11a &0:42:42.81&2.34&3315.3 &  	   &   &5925 (HRC-I)&31&$<1.4$& \\
         &41:18:27.8&&        & 	   &   &5926m (HRC-I)&52&$<1.1$& \\
	 &          &&        & 	   &   &5927m (HRC-I)&84&$<2.5$& \\
	 &          &&        & 	   &   &5928 (HRC-I)&108&$<1.9$& \\
\noalign{\medskip}
2004-11b &0:43:07.45&4.76&3315.3 &  	  &   &5925 (HRC-I)&31&$<3.5$& \\
         &41:18:04.6&&       &  	   &   &5926m (HRC-I)&52&$<4.2$& \\
	 &          &&       &  	   &   &5927m (HRC-I)&84&$11.5\pm1.4$& \\
	 &          &&       &J004307.4+411804&0.1&5928 (HRC-I)&108&$16.2\pm1.6$& \\
\noalign{\medskip}
2004-11g &0:42:52.48&2.41&3315.3 &J004252.4+411800&0.3&5925 (HRC-I)&31&$27.5\pm1.8$& SSS (ACIS-I)\\
         &41:18:00.2&&        & 	 &   &5926m (HRC-I)&52&$82.1\pm3.0$& \\
	 &          &&        & 	   &   &5927m (HRC-I)&84&$5.9\pm0.9$& \\
	 &          &&        & 	   &   &5928 (HRC-I)&108&$2.6\pm0.6$& \\
\noalign{\medskip}
2004-11c &0:42:32.29&3.99&3326.4  & 	   &   &5925 (HRC-I)&20&$<1.8$& \\
         &41:19:25.7&&        & 	   &   &5926m (HRC-I)&41&$<1.7$& \\
	 &          &&        & 	   &   &5927m (HRC-I)&73&$<2.2$& \\
	 &          &&        & 	   &   &5928 (HRC-I)&99&$<2.3$& \\
\noalign{\smallskip}
\hline
\noalign{\smallskip}
\end{tabular}
\label{tab:novae_new1}
\normalsize
\end{table*}

\begin{table*}
\addtocounter{table}{-1}
\caption[]{continued. 
	   }
\scriptsize
\begin{tabular}{lrrrlrrrrl}
\hline\noalign{\smallskip}
\hline\noalign{\smallskip}
\multicolumn{4}{l}{\normalsize{Optical nova candidate}} 
& \multicolumn{4}{l}{\normalsize{X-ray measurements}} \\
\noalign{\smallskip}\hline\noalign{\smallskip}
\multicolumn{1}{l}{Name} & \multicolumn{1}{c}{RA~~~(h:m:s)$^a$} 
& \multicolumn{1}{c}{Offset$^b$} 
& \multicolumn{1}{c}{JD$^c$} & \multicolumn{1}{l}{Source name$^d$}
& \multicolumn{1}{c}{$D$} 
& \multicolumn{1}{c}{Observation$^e$} 
& \multicolumn{1}{c}{$\Delta T^f$} & \multicolumn{1}{c}{$L_{\rm X}^g$}
& \multicolumn{1}{l}{Comment} \\
M31N & \multicolumn{1}{c}{Dec~(d:m:s)$^a$} &\multicolumn{1}{c}{(\arcmin)}
& \multicolumn{1}{l}{2\,450\,000+} & &(\arcsec)  
& \multicolumn{1}{c}{ID} &\multicolumn{1}{c}{(d)} \\
\noalign{\smallskip}\hline\noalign{\smallskip}
2004-11d &0:42:45.46&0.46&3334.2  & 	   &   &5925 (HRC-I)&12&$<0.9$& \\
         &41:16:33.2&&        & 	   &   &5926m (HRC-I)&33&$<1.0$& \\
	 &          &&        & 	   &   &5927m (HRC-I)&65&$<1.4$& \\
	 &          &&        & 	   &   &5928 (HRC-I)&89&$<1.3$& \\
\noalign{\medskip}
2004-11e &0:43:31.85&11.01&3339.3  & 	   &   &5925 (HRC-I)&7&$<16.3$& \\
         &41:09:42.6&&        & 	   &   &5926m (HRC-I)&28&$<9.9$& \\
	 &          &&        &J004331.6+410943&1.9&5927m (HRC-I)&60&$35.9\pm4.0$& \\
	 &          &&        & 	   &   &5928 (HRC-I)&84&$9.1\pm3.5$& \\
\noalign{\medskip}
2005-01a &0:42:28.39&3.03&3378.3$^*$  & 	   &   &5926m (HRC-I)&-11&$<1.3$& \\
         &41:16:36.1&&        & 	   &   &5927m (HRC-I)&21&$<1.3$& \\
	 &          &&        & 	   &   &5928 (HRC-I)&45&$<1.6$& \\
\noalign{\medskip}
2005-02a &0:42:52.79&2.30&3420.3  & 	   &   &5927m (HRC-I)&-21&$<1.5$& \\
         &41:14:28.9&&        & 	   &   &5928 (HRC-I)&3&$<1.5$& \\
\noalign{\smallskip}
\hline
\noalign{\smallskip}
\end{tabular}

\label{tab:novae_new2}
Notes: \hspace{0.3cm} $^a$: RA, Dec are given in J2000.0 \\
\hspace*{1.cm} $^b$: projected distance from \m31\ nucleus position
\citep[RA$_\mathrm{J2000} = 00^h42^m44\fs324, \delta_\mathrm{J2000} =
+41\degr16\arcmin08\farcs53$; ][]{1992ApJ...390L...9C}\\
\hspace*{1.cm} $^c$: $^*$ indicates well defined start date of optical outburst, 
$^<$ outburst start before, else badly defined (see text) \\
\hspace*{1.cm} $^d$: full source names are CXOM31~Jhhmmss.s+ddmmss \\
\hspace*{1.cm} $^e$: for \xmm\  ``mrg" indicate the merged EPIC data of 
ObsID 0202230201, 0202230401, and 0202230501\\ 
\hspace*{1.3cm}for \chandra\ we give ObsID and camera used (see text). 5926m
 indicate the merged data of ObsIDs 6177 and 5926, 5927m those of 6202 and 5927\\
\hspace*{1.cm} $^f$: time after observed start of optical outburst\\
\hspace*{1.cm} $^g$: un-absorbed luminosity in 0.2--1.0 keV band assuming a 50 eV 
black body spectrum with Galactic absorption, upper limits are 3$\sigma$ \\
\normalsize
\end{table*}

\subsubsection{Nova M31N~2003-11a}
This nova candidate was first reported by \citet{2003IAUC.8248....2H}. 
\citet{2003IAUC.8253....3F} constrained the date of outburst and WeCAPP
(N2003-12) provided accurate positioning. 
In X-rays a counterpart was first detected  in the 
\chandra\ HRC-I observation 5925 about 400
days after the optical outburst. The source is detected
throughout HRC-I observation 5928. The upper limit from
the \chandra\ ACIS-I observation 4723 leads to an ACIS-I/HRC-I  
count rate factor less than 0.6. This clearly indicates that the source had a 
supersoft spectrum.

\subsubsection{Nova M31N~2003-11b}
This nova candidate was first reported by \citet{2003IAUC.8253....3F}.
\citet{2004IAUC.8262....2M} constrained the date of outburst and WeCAPP
(N2003-13) provided accurate positioning. 
In X-rays a counterpart was first detected  in the 
\chandra\ HRC-I observation 5925 about 370
days after the optical outburst. The source is detected
throughout HRC-I observation 5928. The upper limit from
the \chandra\ ACIS-I observation 4723 leads to an ACIS-I/HRC-I  
count rate factor less than 0.9. This indicates that the source had a 
soft spectrum.

\subsubsection{Nova M31N~2004-05b}
This nova candidate was detected by Hornoch and Ku\v{s}nir\'ak (see Appendix B, No. 13)
after the \m31\ visibility window opened again in May 2004.
Therefore the time of optical outburst is not well defined. 
In X-rays a counterpart was first detected  in the 
\chandra\ HRC-I observation 5926m about 220
days after the optical outburst with a linear  increase in brightness during 
the following two observations (till day 280 after optical outburst). The source
is not detected during \chandra\ HRC-I observation 5925. We therefore can not
constrain the X-ray spectrum using the ACIS-I/HRC-I count rate ratios. 

\subsubsection{Nova M31N~2004-06a}
This nova candidate was detected by Hornoch (see Appendix B, No. 15). The optical outburst 
could have happened at most 8 days before detection. 
In X-rays a source
coincident with the nova position was first detected 87 days after the optical
outburst. The X-rays reached a maximum around day 120 when the source was
detected as bright SSS (black body temperature around 70 eV, see below). About
200 days after outburst the SSS phase ended.

\begin{figure}
   \resizebox{\hsize}{!}{\begin{minipage}[b]{9cm}
     \includegraphics[height=9cm,angle=90,clip]{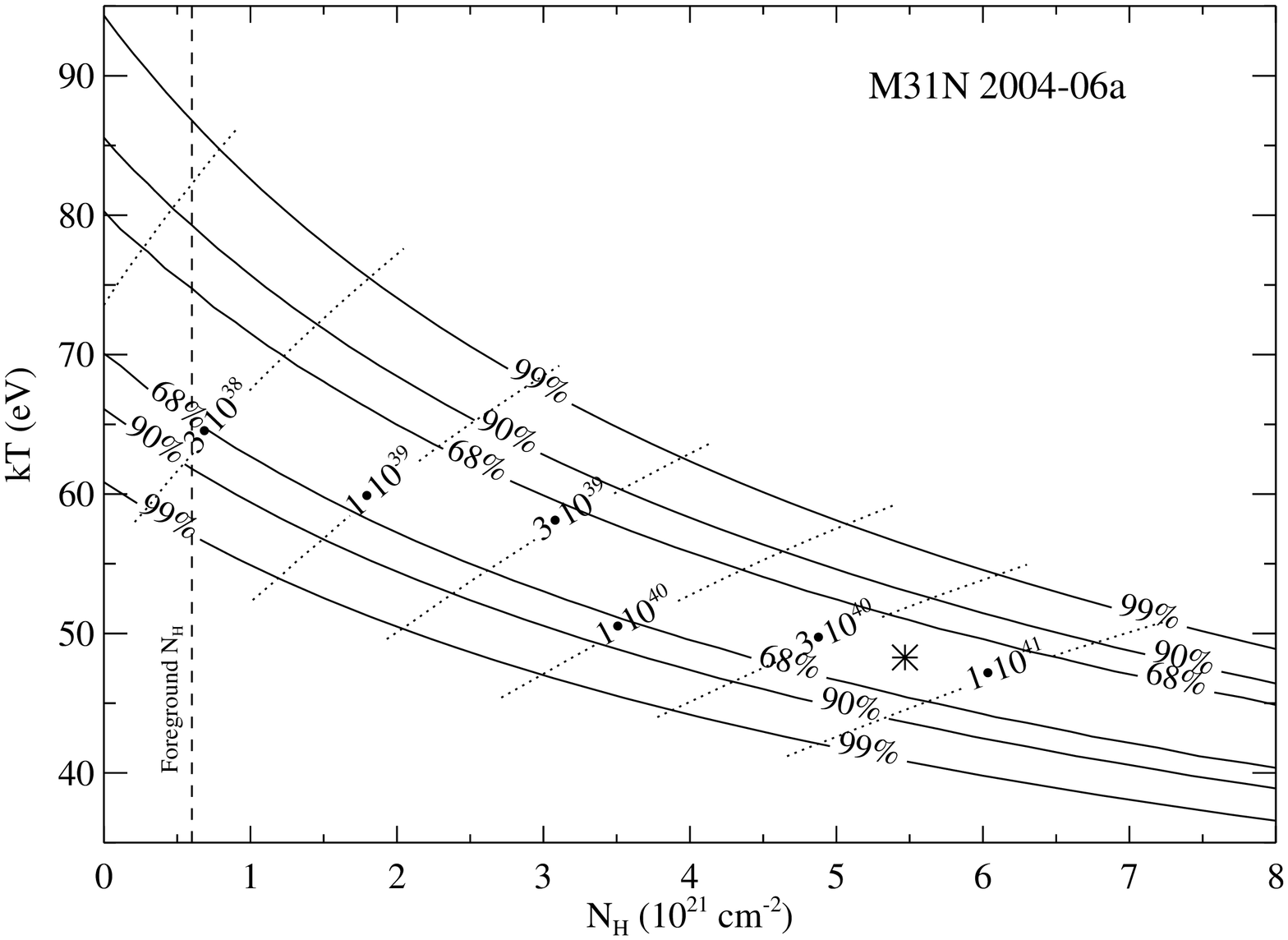}
     \includegraphics[height=9cm,angle=90,clip]{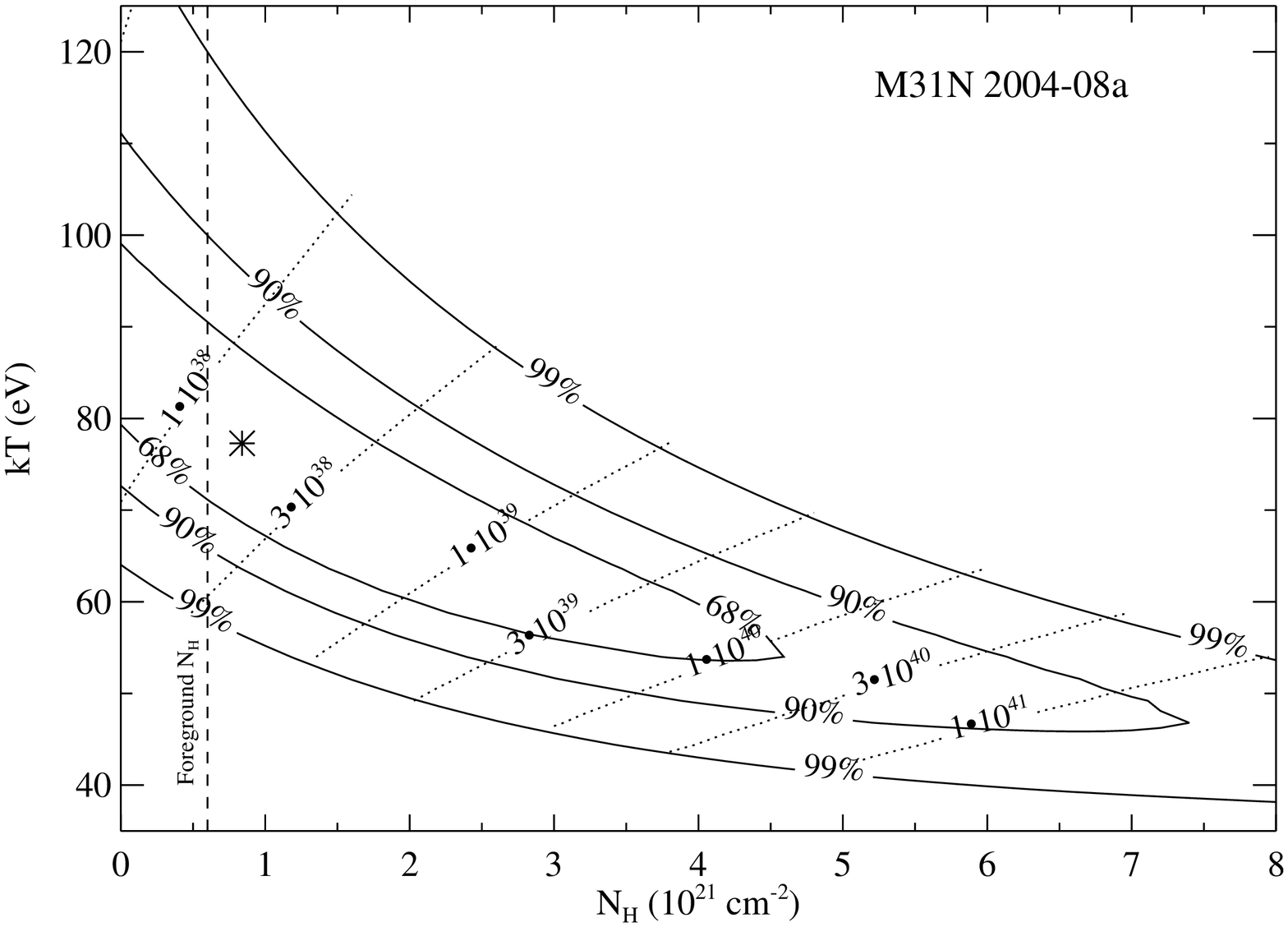}
     \end{minipage}}
     \caption[]{ Column density -- temperature confidence contours inferred from the fit to the ACIS-I 
               spectra of M31N~2004-06a obtained from \chandra\ ACIS-I 
	       observations 4721 and 4722 (above) and of M31N~2004-08a obtained 
	       from \chandra\ ACIS-I observations 4721 (below). The formal best 
	       fit parameters are indicated by stars. Also drawn are lines of 
	       constant bolometric luminosity and for the Galactic foreground 
	       absorption (see Fig.~\ref{pncont_2000_07a}).
               }
         \label{fig:spec_cont}
   \end{figure}

\chandra\ ACIS-I spectra were obtained for this nova from observations 4721 and
4722. Although in total only 84 counts were collected, their distribution at
energies below $\sim$800 eV allows to derive constraints from the X-ray
spectra. The spectra were  binned to contain at least 10 counts per bin and
simultaneously fit by an absorbed 
\citep[tbabs in XSPEC,][]{2000ApJ...542..914W} blackbody
model. Confidence contours for absorption  column density and blackbody
temperature are shown in Fig.~\ref{fig:spec_cont}.
 
\subsubsection{Nova M31N~2004-06c}
This nova candidate was detected by Hornoch (see Appendix B, No. 17). The start of the 
optical outburst is only defined to about 2 weeks. 
In X-rays a counterpart was detected  during the \chandra\ HRC-I observations
5925 to 5928. The X-ray intensity decreased linearly by a factor 3 to 4 from day
165 to 242. The upper limit from
the \chandra\ ACIS-I observation 4723 leads to an ACIS-I/HRC-I  
count rate factor less than 0.4. This clearly indicates that the source had a 
supersoft spectrum.

\subsubsection{Nova M31N~2004-08a}
This candidate for a fast nova was detected by Hornoch and \v{S}arounov\'a (see Appendix B, No. 19).
The optical outburst could have happened at most 3 days before detection.  
In X-rays a counterpart was only significantly detected 64 days after the optical
outburst during \chandra\ ACIS-I observation 4721. 
A \chandra\ ACIS-I spectrum with 42 counts was accumulated from this
observation. 
The analysis was performed similar to the case of M31N~2004-06a and the NH-kT
contours are plotted in Fig.~\ref{fig:spec_cont}. The spectrum indicates a SSS
with a black body spectrum of 80 eV or less. The source was not visible in
ACIS-I observations 30 days earlier or later.

\subsubsection{Nova M31N~2004-08c}
This nova candidate was first reported by \citet{2004ATel..330....1T}. The time of the
optical outburst is not well defined. WeCAPP (N2004-04) provided
accurate positioning.
In X-rays a counterpart was detected 107 days after the optical outburst in the 
\chandra\ HRC-I observation 5925 and was barely visible 20 days later. 
The upper limit from the \chandra\ ACIS-I observation 4723 leads to an ACIS-I/HRC-I 
count rate factor less than 0.4. This clearly indicates that the source had a 
supersoft spectrum.
 
\subsubsection{Nova M31N~2004-11f}
The optical detection of this nova candidate was discussed in Section
\ref{sect:opt}.  The optical outburst 
could have happened at most 7 days before the HST detection.
In X-rays a counterpart was detected 34 days after the optical outburst in the 
\chandra\ ACIS-I observation 4723. 
The count rate from the \chandra\ ACIS-I observation 4723 leads to an ACIS-I/HRC-I
count rate factor of 0.07. This clearly indicates that the source had a 
supersoft spectrum. The source intensity dropped within 20 days by a factor of
more than 30. The X-ray source was no longer detected after 90 days.

Following the discussion in Section \ref{sect:opt} the nova may well be a 
recurrent nova ($\Delta T \sim 20.3$ yr).

\subsubsection{Nova M31N~2004-11b}
This nova candidate was detected by Hornoch (see Appendix B, No. 24).
The optical outburst could have happened at most 11 days before detection
taking into account the last WeCAPP non detection before the Hornoch discovery.  
WeCAPP (N2004-09) in addition provided accurate positioning. Filippenko et al.
\footnote{see http://cfa-www.harvard.edu/iau/CBAT\_M31.html} 
spectroscopically confirmed the variable as nova in \m31.
In X-rays a counterpart was detected 84 days after the optical outburst in the 
\chandra\ HRC-I observations 5927m and increased in brightness by a factor of
1.5 within the 24 days to observation 5928. The source
is not detected during \chandra\ HRC-I observation 5925. We therefore can not
constrain the X-ray spectrum using the ACIS-I/HRC-I count rate ratios. 

\subsubsection{Nova M31N~2004-11g}
The optical detection of this nova candidate was discussed in Section
\ref{sect:opt}.  The optical outburst 
could have happened at most 11 days before the detection.
In X-rays a counterpart was detected 31 days after the optical outburst in the 
\chandra\ HRC-I observation 5925, increased in brightness by a factor of
$\sim3$ to observation 5926m 21 days later. After a drop in intensity by a
factor of more than 10 within 32 days it was barely visible in observations
5927m and 5928. The upper limit from the \chandra\ ACIS-I observation 4723 
leads to an ACIS-I/HRC-I  
count rate factor less than 0.2. This clearly indicates that the source had a 
supersoft spectrum.

\subsubsection{Nova M31N~2004-11e}
This nova candidate was detected by Hornoch (see Appendix B, No. 27) after a gap in
observations of 18 days.   
In X-rays a counterpart was detected 31 days after the optical outburst in the 
\chandra\ HRC-I observation 5927m. It had already dropped in
intensity 24 days later by a factor of more than 3. The source
is not detected during \chandra\ HRC-I observation 5925. We therefore can not
constrain the X-ray spectrum using the ACIS-I/HRC-I count rate ratios.

\section{Discussion and conclusions}
\begin{table*}
\begin{center}
\caption[]{Observed and derived Parameters of X-ray detected optical novae in
\m31.}
\begin{tabular}{lrlrrrrrrr}
\hline\noalign{\smallskip}
\hline\noalign{\smallskip}
\multicolumn{4}{l}{\normalsize{Optical measurements}} 
& \multicolumn{4}{l}{\normalsize{X-ray measurements}} 
& \multicolumn{2}{l}{\normalsize{derived parameters}}\\
\noalign{\smallskip}\hline\noalign{\smallskip}
\multicolumn{1}{c}{Name$^a$} &\multicolumn{1}{c}{R$^b$} &
\multicolumn{1}{c}{Brightness$^c$} &
\multicolumn{1}{c}{$t_{\rm 2R}$} &
\multicolumn{2}{c}{SSS phase} &
\multicolumn{1}{c}{$L_{\rm X}^{d}$}  &  \multicolumn{1}{c}{$kT_{\rm BB}^e$} &
\multicolumn{1}{c}{Ejected mass} & \multicolumn{1}{c}{Burned mass} \\
\noalign{\smallskip}
M31N& &(mag Filter)& (d)  
& \multicolumn{1}{c}{Start (d)} 
& \multicolumn{1}{c}{Turn-off (d)}
& \multicolumn{1}{c}{}& \multicolumn{1}{c}{(eV)}
& \multicolumn{1}{c}{($10^{-5}$ \msun)} & \multicolumn{1}{c}{($10^{-6}$ \msun)}\\
\noalign{\smallskip}\hline\noalign{\smallskip}
1995-09b & & 15.6 H$\alpha$ &  &  $ - $       & $2327-3383$	& 16.1  & & $-$ 	  & $3.9-5.7$	  \\
1995-11c & & 16.3 H$\alpha$ &  &  $ - $       & $>3373$ 	& 13.8  & & $-$ 	  & $>5.7$	  \\
1996-08b & r& 16.1 H$\alpha$ &  &  $1782-1880$ & $>3115$ 	& 5.6	& & $330-370$	  & $>5.2$	  \\
1997-11a & r& 18.0 R	    &  &  $1461-2593$ & $>2670  $	& 4.4	& & $220-700$	  & $>4.5$	  \\
1998-06a & r& 16.3 H$\alpha$ &  &  $-$	      & $1310-2235  $	& 1.7	& & $-$ 	  & $2.2-3.7$	  \\
1999-10a & & 17.5 w	    &  &  $760-1751$  & $>1969 $	& 21.2  & $30-38$& $61-320$	  & $>3.3$	  \\
2000-08a & & 18.6 R	    &  &  $203-253$   & $494-1627$	& 16.8  & & $4.3-6.7$	  & $0.83-2.7$    \\
2000-07a$^*$ & & 16.8 R     & 22.4 &   $154-170$   & $>1668$  	& 13.5 	& $28-37$& $2.5-3.0$	  & $>2.8$	  \\
2001-08d$^*$ & & 16.7 R     & 11.8 &   $<63$       & $<1055$    &  0.7  & & $<0.42$	  & $<1.8$	  \\
2001-10a$^{*+}$ & & 17.0 R     & 39.3 &  $1019-1158$ & $>1235$  &  5.2  & & $110-140$	  & $>2.1$  \\ 
2001-10f & & 16.6 B         &  &   $17-84$     & $<1009$        & 37.0 	& & $0.03-0.74$  	& $<1.7$ \\
2003-11a$^*$ & & 16.9 R     &  &  $256-396$   & $>473$   	& 27.6  & & $6.9-16$  	& $>0.79$  \\
2003-11b$^*$ & & 17.4 R     & 42.2 &   $227-367$   & $>444$   	& 10.4 	& & $5.4-14$  	& $>0.75$  \\
2004-05b & & 17.2 R         & 49.7 &   $202-223$   & $>279 $   	& 19.4  & & $4.3-5.2$  	& $>0.47$  \\
2004-06a & & 17.2 R         & 19.7 &   $41-87$     & $145-180$  & 62.1  & $63-87$& $0.2-0.8$  	& $0.24-0.30$  \\
2004-06c & & 17.1 R         & 10.9 &   $24-165$   & $>242$      & 33.9  & & $0.6-2.9$  	& $>0.41$  \\
2004-08a & & 17.4 R         &  &    $32-64$   & $64-90$         & 51.9  & $60-120$& $0.1-0.4$  	& $0.11-0.15$  \\
2004-08c & & 18.7 R         & 50.3 &       $<107$    & $128-160$   	&  9.3  & & $<1.2$  	& $0.21-0.27$  \\
2004-11f & ?& 17.9 R         & 28.4 &       $<34$     & $35-55 $   	& 353.3 & & $<0.1$     	& $0.06-0.09$  \\
2004-11b$^+$ & & 16.6 R         & 32.0 &  $52-84$   & $>108 $   	& 16.2  & & $0.3-0.7$	& $>0.18$  \\
2004-11g & & 17.9 R         & 28.4 &       $<31$    	& $52-84$   	& 82.1  & & $<0.1$	& $0.09-0.14$  \\
2004-11e & & 17.6 R         & 34.6 &   $28-60$   & $>84 $    	& 35.9  & & $0.08-0.4$	& $>0.14$  \\
\noalign{\smallskip}
\hline
\noalign{\smallskip}
\end{tabular}
\label{tab:novapar}
\end{center}
Notes:\hspace{0.3cm} $^a $: $^*$ indicates that the date of outburst is well 
defined, $^+$  indicates novae confirmed by optical spectra\\
\hspace*{1.1cm} $^b $: Flag for recurrent novae: ``r" indicates recurrent nova candidates, ``?" the
recurrent candidate discussed in Sect.~\ref{sect:opt}\\
\hspace*{1.1cm} $^c $: ``w" indicates without filter \\
\hspace*{1.1cm} $^d $: un-absorbed luminosity in 0.2--1.0 keV band in units of 
10$^{36}$ erg s$^{-1}$ during observed maximum X-ray brightness assuming \\
\hspace*{1.5cm} a 50 eV black body spectrum with Galactic absorption\\
\hspace*{1.1cm} $^e $: allowed black body temperature range from spectral fit \\
\end{table*}

Optical novae have recently been identified as the major class of supersoft
X-ray sources in M~31 based on ROSAT and early \xmm\ and \chandra\ 
observations (PFF2005). 
In this second paper on our program to systematically search for X-ray emission
from optical novae in \m31\ we concentrate on archival \chandra\ observations to
the \m31\ center. While in the first paper the sampling was mainly based on four
\xmm\ observations separated by half a year (together with ROSAT
and some \chandra\ observations), we here concentrate on four HRC-I observations 
with a 20 to 30 day spacing obtained from December 2004 to February 2005. 
Five ACIS-I observations with a monthly spacing (July to  December 2004), the 
last one 1.7 days before the time of the first 
HRC-I observation, were of great importance to extend the time coverage, even 
when the ACIS-I CCDs are
only sensitive to detect extremely bright nova counterparts with a supersoft spectrum 
or counterparts that have spectra at the high temperature end for SSS. 
We also included in the analysis four \xmm\ 
observations of the \m31\ center region
obtained within four days in July 2004.
In this way we were able to cover a large enough time base 
to detect not only several nova 
counterparts with long duration SSS phase but also 
a significant number of nova counterparts with short SSS
phase (SSS turn-off $<$100~d after outburst). 

For comparison with the X-ray data we created a catalogue of optical 
novae in M~31 based on our own nova search programs and on all novae 
reported in the literature. We collected all known properties and named the 
novae consistently following the CBAT scheme.

Some nova counterparts are still detected in X-rays several thousand days after 
the optical nova outburst.
The three extremes are M31N~1995-09b, M31N~1995-11c and
M31N~1996-08b detected 
3383, 3373 and 3115 days (i.e. 9.3, 9.2 and 8.5 years) after outburst, 
respectively. This can be compared to
nova counterparts in the Galaxy and Magellanic Clouds with long SSS states: 
nova V723 Cas detected as SSS more than 11 years after outburst
\citep{2006CBET..598....1N};
GQ Mus, SSS for less than 9 years \citep{1993Natur.361..331O};
nova LMC 1995, SSS for less than 8 years 
\citep[see ][]{2003ApJ...594..435O,2004RMxAC..20..182O}. 

Inspecting Fig.~\ref{fig:LCs}, two novae show a linear rise in X-ray flux
(M31N~2004-05b, M31N~2004-11b) and one a linear decay 
(M31N~2004-06c). However,
more resolved X-ray light curves are needed to see if this behavior is common
for many novae and deserves further modeling. 
One nova counterpart is detected to turn on as SSS
50~d, another 200~d after the optical nova outburst. Three nova counterparts 
unexpectedly showed short X-ray outbursts starting
within 50~d after the optical outburst and lasting only two to three months.
From Fig.~\ref{fig:LCs} it is clear that specifically for novae with
short SSS states, the observations in the 
archive are not sampling the light curves with sufficient resolution and
sensitivity to really allow us to resolve the outbursts for detailed modeling. 
For such a task, dedicated observations are urgently needed.
Nevertheless, many interesting results can already be deduced from the available
observations. 

In Table~\ref{tab:novapar} we combine optical and X-ray properties of 
the X-ray detected novae presented in Tables~\ref{tab:novae_old} and 
\ref{tab:novae_new} and as derived parameters the ejected and burned masses. 
The optical information includes accuracy with which the outburst date is
known, nova confirmation by optical spectra, a  flag for recurrent novae, 
observed brightness in outburst and the fastness parameter $t_{\rm 2R}$.
From the X-ray measurements we give start of SSS phase, turn-off of SSS phase 
(defined as factor of 10
decrease from maximum X-ray luminosity), the maximum X-ray luminosity and the
allowed (3$\sigma$) temperature range from a black body fit.
M31N~1994-09a, M31N~1997-08b and  M31N~2002-01b 
are not
included as start and turn-off time of the SSS state are not well determined and
no useful mass limits can be given.

The un-absorbed X-ray luminosity in the 0.2--1.0 keV band during maximum is
rather uncertain and the tabulated values have to be taken with care. 
For nova counterparts with a short SSS phase, the time of
maximum X-ray brightness may not be covered by the observations. Also, 
X-ray brightnesses calculated assuming a 50 eV 
black body spectrum with Galactic absorption are very uncertain. 
As the spectral fits to
several novae show, the surface temperature can vary significantly from nova to
nova (see discussion below) which would alter the un-absorbed X-ray luminosity by
factors of 10 or more. Also additional absorption within \m31\ may
enhance the computed un-absorbed X-ray luminosities. 

The ejected mass in the nova outburst can be approximately determined from the
start date of the SSS phase. The decrease of the optical thickness of the
expanding ejecta is responsible for the rise in the X-ray  light curve of the
post-outburst novae, as shown for V1974 Cyg 
\citep{1996ApJ...463L..21S,1996ApJ...456..788K}. Assuming the material ejected
by the nova explosion to form a  spherical homogeneous shell expanding at
constant velocity $v$,  the hydrogen mass density of the shell will evolve in
time $t$ like $\rho =\frac{M^{ej}_{H}}{\frac{4}{3}\pi v^{3}t^{3}}$ where
$M^{ej}_{H}$ is the ejected  hydrogen mass \citep{1996ApJ...456..788K}.
Assuming an homogeneous density, the column density of hydrogen will evolve
with time like  $N_{H} ({\rm cm}^{-2})=\frac{M^{ej}_{H}}{\frac{4}{3}\pi
m_{H}v^{2}t^{2}}$, where $m_{H}=1.673\times10^{-24}$ g is the mass of the
hydrogen atom.   Assuming typical values for the expansion velocity (2000 km
s$^{-1}$)  and that the SSS turns on when the absorbing hydrogen column density
decreases  to $\sim10^{21}$ cm$^{-2}$, we determine the ejected mass for each
nova in  Table~\ref{tab:novapar}. Typical ejected masses are up to a few times 
$10^{-5}$ \msun. These estimates can be improved for novae with measured
outflow velocities. Unfortunately, only for one nova in our sample an
expansion  velocity was derived from an optical spectrum  (2500 km s$^{-1}$ for
M31N~2004-11b)\footnote{see Filippenko et al. at \\
http://cfa-www.harvard.edu/iau/CBAT\_M31.html}.  For the novae that have a SSS
phase starting more than one year after the  optical outburst,  the ejected
masses are unrealistically large.  Some of them are proven recurrent novae
(M31N~1996-08b,  M31N~1997-11a; see Table A.1), and thus the 
ejected mass calculated from the SSS turn-on time is meaningless.  In the other
cases of delayed SSS, it  may be RN or SSS phases that started again on the hot
post-novae WD surface without an optical outburst \citep[see
][]{2004ApJ...612L..53S}. For some of the novae we can not constrain the start
times of the SSS state in a useful manner and consequently no ejected  mass is
estimated. 
 
The turn-off time of the SSS phase indicates directly the amount of hydrogen rich 
material burned on the 
WD surface, $M^{burn}={\frac{L\Delta t}{X_H \epsilon}}$, where $L$ is 
the bolometric luminosity,
$\Delta t$ is the duration of the SSS phase, $X_H$ is the hydrogen fraction of 
the burned material and 
$\epsilon=5.98\times10^{18}$ erg g$^{-1}$. We compute the burned mass during the  
SSS phase for each novae, assuming a bolometric luminosity of 
$3\times10^4L_{\odot}$ and a hydrogen mass fraction 
$X_H=0.5$. The values given in Table~\ref{tab:novapar} correspond to the mass 
burned assuming that the hydrogen burning phase of a nova starts with the time 
of outburst and ends with the turn-off of the SSS phase. For most of the novae
the burned mass is below a few times $10^{-6}$ \msun. For novae in which the SSS
phase has not ended during our observations we only give lower limits for the
burned mass. However, none of them is extraordinarily high. If in some novae the 
SSS phase (and also the hydrogen burning phase) should be delayed as proposed
above, the burned
mass estimates would just represent upper limits.

In spite of the uncertainties and the fact that for most novae we 
only obtain upper or lower limits, for the cases with constrained burned 
and ejected mass (i.e., novae M31N~2000-08a, M31N~2004-06a 
and M31N~2004-08a) the burned 
mass is about one order of magnitude smaller than the ejected mass. The 
burned masses are within the values expected from models of post-outburst 
novae with stable hydrogen burning envelopes 
\citep{2005A&A...439.1061S,1998ApJ...503..381T}, while the ejected masses 
derived for \m31\ novae are also in the ranges of the ejected masses 
predicted from hydrodynamical models of nova outbursts 
\citep{1998ApJ...494..680J}. But without any information on the accreted 
mass and the degree of mixing of the accreted envelope with the degenerate 
core, we cannot draw a conclusion on the actual increase or decrease of 
the white dwarf mass and its long-term evolution.

As mentioned in Section \ref{sect:novae_new}, 21 of the 32 novae with optical 
outburst from November 2003 to February
2005 have not been detected in X-rays. This could have several reasons: 
(i) the novae could have very short SSS phases which are not covered by our X-ray
sampling, 
(ii) the SSS phase of the novae ended before the X-ray observations in 2004, 
(iii) the SSS phase will only start after the  X-ray observations in 2005,
(iv) last but not least, the novae could not go through a SSS phase.
From the above numbers it is however clear that more than 30\% of the novae 
(and probably even many more) go through a X-ray SSS phase. This percentage 
is about a factor of two above the lower limit given by PFF2005 and corroborates
their expectation that the fraction of novae with SSS phase may be significantly
higher. 

X-ray emission of eight of the 14 newly detected optical nova counterparts 
can be characterized as
supersoft from hardness ratios and spectra or by comparing HRC-I
count rates with ACIS-I count rates or upper limits.
For four nova counterparts we were able to model the X-ray spectrum in greater
detail. While the spectra of novae with longer SSS states have lower
temperatures (M31N~2000-07a and M31N~1999-10a show a $kT$ of 
about 30--35 eV), the 
novae with short SSS state seem to have significantly higher temperatures
(M31N~2004-06a and M31N~2004-08a about 70--80 eV). This fits to the nova 
outburst behavior calculated by \citet{2006ApJS..167...59H} that was
successfully applied to predict and model the X-ray light curve of the recurrent
nova RS Oph \citep{2006ApJ...642L..53H,2006ApJ...651L.141H}, V1974 Cyg 
\citep{2005ApJ...631.1094H}, V693 CrA, V1668 Cyg, V351 Pup and OS And 
\citep{2006astro.ph.11594K}. In these
models novae with lower mass WDs show longer lasting SSS states with
lower temperatures,  novae with higher mass WDs show shorter SSS states with
higher temperatures.

Based on these and earlier results we initiated an optical (photometry and
spectroscopy) and X-ray monitoring program for novae in the central region of 
\m31. In this way we hope to get a better handle on the percentage of novae
showing SSS states and novae with short SSS states.
 
\begin{acknowledgements}
This work uses data from the archive of the Isaac Newton Group of
telescopes, maintained at the Institute of Astronomy, Cambridge; 
it also used information provided on the ``M31 (Apparent) Novae Page"
provided by the International Astronomical Union, Central Bureau for 
Astronomical Telegrams CBAT 
(http://cfa-www.harvard.edu/iau/CBAT\_M31.html) and the finding
charts and  information, collected by David Bishop 
(``Extragalactic Novae - 2003", 
http://www.rochesterastronomy.org/sn2003/novae.html and following years).
This research has made use of the VizieR catalogue access tool, CDS,
Strasbourg, France. We are grateful to Daniel Green from CBAT who encouraged us 
``to use the CBAT 
designation scheme for novae in \m31\ and to extend it to all novae in \m31".
We want to thank the editor, Steven N. Shore, for his comments, which
helped to improve the manuscript considerably.
The optical nova search in \m31\ was supported by the Grant Agency of the Czech
Republic, grant No. 205/04/2063. The WeCAPP program was supported by the 
\emph{Son\-der\-for\-schungs\-be\-reich, SFB\/} 375
of the \emph{Deut\-sche For\-schungs\-ge\-mein\-schaft, DFG\/}. 
We are grateful for spectroscopic confirmation of nova candidates by 
A. V. Filippenko,
R. J. Foley, M. Ganeshalingam, F. J. D. Serduke, J. L. Hoffman,
S. Jha, M. Papenkova (of novae M31N~2002-08a, M31N~2004-08b, 
M31N~2004-09a, M31N~2004-11b, M31N~2004-11a, 
M31N~2005-01a, and M31N~2005-07a),
to F. Di Mille, S. Ciroi, V. Botte, C. S. Boschetti (of novae 
M31N~2003-08c and M31N~2003-09b)
and to M. Della Valle, A. Pizzella, E. Giro, N. Hripsime, L. Coccato, and
I. Iegoriva (of nova M31N~2005-01a).
The X-ray work is
based in part on observations with \xmm, an ESA Science Mission 
with instruments and contributions directly funded by ESA Member
States and NASA.
The \xmm\ project is supported by the Bundesministerium f\"{u}r
Wirtschaft und Technologie / Deutsches Zentrum f\"{u}r Luft- und Raumfahrt 
(BMWI/DLR FKZ 50 OX 0001), the Max-Planck Society and the Heidenhain-Stiftung.
G.S. is supported by a postdoctoral fellowship of the Spanish Ministry 
for Education and Science.

\end{acknowledgements}
\bibliographystyle{aa}
\bibliography{./novae_2004,/home/wnp/data1/papers/my1990,/home/wnp/data1/papers/my2000,/home/wnp/data1/papers/my2001,/home/wnp/data1/papers/my2004,/home/wnp/data1/papers/my2006}

\appendix
\section{Master catalog of novae detected in \m31}
For cross-correlating with X-ray sources and searching for recurrent novae we
created a catalogue of all historical optical novae detected in \m31.
During this work we noticed that a homogeneous naming of all novae in \m31\ is
missing. After discussions with Nicolai Samus, who's group provided naming for
some novae in the "General Catalogue of Variable Stars" 
\citep{2004yCat.2250....0S} and Daniel Green from the Central Bureau for 
Astronomical Telegrams CBAT of the International Astronomical Union, we decided
to extend the CBAT naming scheme to all optical novae and candidates in \m31\ 
that have been reported in the literature and make the information available 
in Table~A.1 (novae with outbursts before end of 2005) 
together with additional parameters and references (Table~A.2)
and in extended form via
the Internet\footnote{see
http://www.mpe.mpg.de/$\sim$m31novae/opt/m31/index.php}. 
We intend to update the Internet pages regularly and encourage observers
to provide input for historical and forthcoming optical novae in \m31. We will
include photometric and spectroscopic data of optical novae and candidates in \m31\
covering all wavelengths. 


We combined the WeCAPP nova list (FBR2007) with novae from other microlensing 
surveys of \m31: the AGAPE survey \citep{2004A&A...421..509A}, 
the POINT-AGAPE PACN survey \citep{2004MNRAS.351.1071A,2004MNRAS.353..571D},  
the Nainital Microlensing Survey \citep{2004A&A...415..471J} and
the survey by \citet{1996AJ....112.2872T}. 
We added novae from IAU circulars, astronomical telegrams (ATel) 
and information bulletins on variable stars (IBVS)
and novae and candidates on the
``M31 (Apparent) Novae Page"
provided by the International Astronomical Union, Central Bureau for 
Astronomical Telegrams CBAT and the finding
charts and  information, collected by David Bishop 
(``Extragalactic Novae - 2003" and following years). We included 
novae from the H$\alpha$ searches of 
\citet{2001ApJ...563..749S},
\citet[][nova and nova candidate lists\footnote{available at
http://www.noao.edu/outreach/rsbe/nova.html}]{1999AAS...195.3608R},
\citet{1992ApJS...81..683T},
\citet{1990ApJ...356..472C}, \citet{1987ApJ...318..520C}
and \citet{1983ApJ...272...92C}. 
We added the lists by Sharov and colleagues 
\citep{1991Ap&SS.180..273S,1992Ap&SS.188..143S,1992Ap&SS.190..119S,
1993AstL...19..230S,1994AstL...20...18S,1994AstL...20..711S,1995AstL...21..579S,
1996AstL...22..680S,1997AstL...23..540S,1998AstL...24..445S,1998AstL...24..641S,
2000AstL...26..433S} and \citet{1989AJ.....97...83R}.

We also included all earlier catalogues:
Rosino and colleagues  
\citep{1973A&AS....9..347R,1964AnAp...27..498R,1958CoAsi..93....1G,
1956CoAsi..77....1G,1955CoAsi..68....1R}, 
\citet{1968AN....291...19B}, \citet{1967AJ.....72.1356M},
\citet{1965AJ.....70..212B}, 
\citet{1962AJ.....67..334G}, \citet{1964ApJ...139.1027B},
\citet{1956AJ.....61...15A}, \citet{1936HDA.....7..671S},
\citet{1931PASP...43..217M}, and of course the pioneering work of 
\citet{1929ApJ....69..103H}.

In the work of \citet{1936HDA.....7..671S}, only the year of the nova outburst is
given.  
\citet{1964ApJ...139.1027B} for one nova explicitly give 1948 as the year of outburst 
for three novae we assumed 1945 as year of outburst. For all these novae for
naming purposes we assumed outbursts on September 1 of the year.
Nova positions in the earlier catalogues (based on optical images obtained with
photographic plates) were often 
only determined to 0.1 arcmin or worse. Including these novae in searches for recurrent novae 
and correlations with X-ray sources is problematic as one might find many
spurious identifications specifically in the 
central region of \m31\ which is crowded with novae and X-ray 
sources. A re-analysis of these archival plates (digitization and position
determination) is partly in progress or urgently should be done. Only then one
will be able to fully exploit these important early data.

The \m31\ nova catalogue in Table~A.1 contains 719 optical novae and nova 
candidates with outbursts before end of 2005. 

\section{Discovery and optical photometry of novae in M~31 by Kamil Hornoch}
An extensive nova search in the center area of \m31\ is performed mostly based
on images obtained with telescopes of the Lelekovice (350/1658 mm)
and Ond\v{r}ejov observatories (650/2342 mm) which are equipped with
a SBIG ST-6V CCD camera with 18\arcmin$\times$13\arcmin\ FOV 
at the Newton focus and an Apogee AP7p CCD camera, 19\arcmin$\times$19\arcmin\ 
FOV at the primary focus, respectively, 
and both mainly use Kron-Cousins R filters.
From Lelekovice, mostly two overlapping fields are imaged if weather and time 
permit, one with the center NE from the \m31\ center and the
second with the center SW from the \m31\ center. Sometimes, 
also regions NW and SE of the center of galaxy are imaged. In almost every
observing night the \m31\ nucleus was covered.
From Ond\v{r}ejov, mostly just one field is imaged centered on the
\m31\ nucleus. In cases, when a nova more distant from the \m31\ center was
discovered,
Ond\v{r}ejov image positions were placed to also cover the nova (up to the 
time when it faded below the sensitivity limit).

Standard reduction procedures for CCD images were applied (bias, dark-frame, 
flat-field correction using the Munipack\footnote{http://munipack.astronomy.cz/} 
or SIMS\footnote{http://ccd.mii.cz/} programs). Reduced images are then 
added. The total
exposure time is typically 10-20 minutes per field (typical exposure times
of partial exposures from Ond\v{r}ejov are 180-sec, from Lelekovice 60-sec).
Gradient of the galaxy background of added images is flattened by median
filter using SIMS. Last procedure is aperture photometry (using 
GAIA\footnote{http://star-www.dur.ac.uk/$\sim$pdraper/gaia/gaia.html}) and astrometry 
(using APHOT, a synthetic aperture photometry and astrometry software developed 
by M. Velen and P. Pravec at the Ond\v{r}ejov Observatory). 
Each suspected nova was checked by an
inspection of all contributing images. The comparison stars were calibrated
using Landolt fields and the ``BVRI CCD photometry
of 361281 objects in the field of M 31" \citep{1992A&AS...96..379M}.
Aperture photometry is made by hand because there is necessity to determine
level of background at the place of the nova.
Typical photometric uncertainties are about 0.1-0.15 mag for 
the brighter novae increasing up to $\sim$0.3 mag close to the detection limit 
(i.e. if the nova is faint or if its position is close to the
\m31\ center or if it is placed at a very non-uniform background).
The values of uncertainties are similar for objects of 18 mag
measured from Lelekovice as for objects about 19 mag when they are observed
from Ond\v{r}ejov.
For astrometry we created (with the help of Miroslav Velen and Petr Pravec) 
a special catalog of fainter objects in the \m31\
field derived from Ond\v{r}ejov images corrected to the world coordinate system 
(WCS) with the help of stars from the UCAC2 catalog \citep{2004AJ....127.3043Z}.
Typically, 100--200 stars from this catalog can be used for WCS determination 
on Lelekovice or Ond\v{r}ejov images. Mean residual of catalog positions are 
0.2"--0.3" and derived nova positions mostly better than 1" and frequently better
than 0.5" as comparison with positions of the same novae from larger telescopes
show. With the help of the astrometric catalog we were able to obtain good 
positions also for images from the \m31\ center area with smaller FOV taken 
with larger ground-based telescopes where WCS registration with standard 
catalogues normally would fail. While observations from larger telescopes
normally are just used to complete Lelekovice and Ond\v{r}ejov light curves to
fainter magnitudes, a few novae are only detected in these images (No. 33 and 
35) or in archival H$\alpha$ images (No. 30 and 31).

Positions of all detected novae are given in Table~\ref{tab:novae_hornoch_pos}.
We also included the pre-discovery observations of M31N~2004-11g as
No.~40 and the observations of M31N~2004-07a as No.~41, that not only 
establish from the improved position that this nova detected by Marco 
Fiashi\footnote{see http://cfa-www.harvard.edu/iau/CBAT\_M31.html} end of July 
2004, is the same as the nova WeCAPP N2004-04 detected on November 5, 2004, but
also identifies the nova by its light curve as a slow nova. We give an internal
number (Col. 1), the nova name following the CBAT nomenclature (2), 
the position (3,4) and the offset in RA/Dec from the nucleus (5,6) as well as 
outburst date and brightness (7,8) and -- if possible -- the time interval 
between maximum brightness and the decay to a brightness two magnitudes below 
($t_{\rm 2R}$) as a nova speed indicator (9). We determined these $t_2$ timescales 
with the algorithm described in Sect. \ref{sect:opt}.
The derived values for novae No. 1--3, 5, and 8--10 
nicely agree with the $t_{\rm 2R}$ values reported by 
\citet{2005ASPC..330..449S} using a different method.
Finally, discoverers of the novae (11) and references (12) are given in
which the discoveries were reported.

\begin{table*}
\caption[]{Nova positions of novae measured by Kamil Hornoch on discovery 
images or on best images available at the time of announcement of discovery.
Eight additional positions for novae No. 24, 25, 32, 33, 37 to 39, and 41 were
measured on deep images with better signal to noise for novae and with better
resolution compared to announcement images. Discoverers and references for 
the individual novae
are given under columns ``Dis." and ``Ref.".  For photometric data of individual novae see
Table~\ref{tab:nova_phot} and \ref{tab:hornoch_obs}.	   }
\scriptsize
\begin{tabular}{llrrrrrrrrl}
\hline\noalign{\smallskip}
\hline\noalign{\smallskip}
\multicolumn{1}{c}{No.} & \multicolumn{1}{c}{Name$^a$} & \multicolumn{1}{c}{RA} &
\multicolumn{1}{c}{Dec} &
\multicolumn{2}{l}{Offset} &
\multicolumn{3}{l}{Outburst} & Dis.$^b$ &Refs.$^c$\\
& M31N & \multicolumn{1}{c}{J2000} & \multicolumn{1}{c}{J2000} &
\multicolumn{1}{c}{RA} & \multicolumn{1}{c}{Dec} &  \multicolumn{1}{c}{JD} &
 \multicolumn{1}{c}{R} & \multicolumn{1}{c}{$t_{\rm 2R}$} \\
& &   \multicolumn{1}{c}{(h:m:s)} &  \multicolumn{1}{c}{(+d:m:s)} & & & 2\,450\,000+ &(mag)
&  \multicolumn{1}{c}{(d)} \\ 
\noalign{\smallskip}\hline\noalign{\smallskip}
1 &2002-08a&00:42:30.92&+41:06:13.1 & 152" W  &  596" S & 2490.523 & 17.05&
51.0 (52$^d$) & (a) & (1,10)\\
2 &2003-06b&00:43:36.17&+41:16:39.4 & 583" E  &   30" N & 2815.507 & 16.7&
7.0--40.6 (28$^d$) & (a) & (2) \\
3 &2003-07b&00:42:15.81&+41:12:00.5 & 318" W  &  252" S & 2835.488 & 16.9&
16.6--23.3 (20$^d$) & (a) &  (3) \\
4 &2003-06d&00:42:41.10&+41:18:32.0 &  36" W  &  144" N & 2840.531 & 18.5&
& (a) &  (3)\\
5 &2003-09b&00:42:46.72&+41:19:46.7 &  26" E  &  219" N & 2913.335 & 17.0&
52.7 (53$^d$) & (a) & (4,11)\\
6 &2003-08c&00:42:41.09&+41:16:16.3 &  37" W  &    8" N & 2930.476 & 17.9&
& (b) & (11)\\
7 &2003-11a&00:42:53.64&+41:18:45.9 &  105" E &  157" N & 2952.573 & 16.9&
& (a) & (5)\\
8 &2003-12a&00:43:04.77&+41:12:23.0 &  231" E &  226" S & 2997.195 & 17.8&
34.2 (35$^d$) & (a) & (6) \\
9 &2004-01a&00:43:08.65&+41:15:35.4 &  274" E &   33" S & 3035.232 & 17.8&
46.5--87.3 (48$^d$) & (a) & \\
10&2004-03a&00:42:36.21&+41:15:37.9 &   92" W &   31" S & 3068.267 & 17.3&
11.0 (12$^d$) & (a) & \\
11&2004-03b&00:43:06.72&+41:11:58.5 &  253" E &  250" S & 3079.272 & 18.1&
& (a) & \\
12&2004-05a&00:42:37.55&+41:10:16.4 &   76" W &  352" S & 3144.560 & 17.0&
19.4--28.2  & (b) & \\
13&2004-05b&00:42:37.04&+41:14:28.5 &   82" W &  100" S & 3143.561 & 17.2&
49.7  & (b) & \\
14&2004-05c&00:43:04.04&+41:23:42.6 &  222" E &  454" N & 3145.563 & 18.4&
36.4  & (b) & \\
15&2004-06a&00:42:22.31&+41:13:44.9 &  248" W &  144" S & 3164.541 & 17.2&
19.7   & (a) & \\
16&2004-06b&00:42:41.30&+41:14:04.2 &   34" W &  124" S & 3178.503 & 17.8&
45.0--47.7  & (a) &  \\
17&2004-06c&00:42:49.02&+41:19:17.8 &   53" E &  189" N & 3181.517 & 17.1&
10.9   & (a) & \\
18&2003-07a&00:42:02.93&+41:05:01.5 &  468" W &  667" S & 2847.480 & 17.6&
$<255.9 $  & (d) & \\
19&2004-08a&00:42:20.62&+41:16:09.5 &  267" W &    1" N & 3220.474 & 17.4&
& (d) & \\
20&2004-08b&00:43:26.84&+41:16:40.8 &  479" E &   32" N & 3225.482 & 17.3&
39.6--63.0  & (b) & (7,12) \\
21&2004-09a&00:42:40.25&+41:14:42.9 &   46" W &   86" S & 3251.518 & 17.5&
23.8-31.5  & (c) &  (7,12)\\
22&2004-10a&00:42:51.84&+41:16:18.2 & 84.7" E &  9.7" N & 3287.835 & 18.6&
& (f) &  \\
23&2004-10b&00:42:47.24&+41:15:54.6 & 32.9" E & 13.9" S & 3288.805 & 18.6&
& (f) & \\
24&2004-11b&00:43:07.43&+41:18:04.4 &260.4" E &115.9" N & 3315.347 & 16.6&
32.0   & (a) & (13)\\
  &        &00:43:07.46&+41:18:04.4  \\
25&2004-11a&00:42:42.76&+41:18:28.0 & 17.6" W &139.5" N & 3315.347 & 16.5&
11.7   & (a) & (13)\\
  &        &00:42:42.80&+41:18:28.0  \\
26&2004-11d&00:42:45.47&+41:16:33.1 & 12.9" E & 24.6" N & 3334.218 & 17.0&
15.3   & (a) & \\
27&2004-11e&00:43:31.85&+41:09:42.6 &536.3" E &385.9" S & 3339.296 & 17.6&
34.6   & (a) & \\
28&2004-12a&00:42:28.05&+41:09:55.6 &183.6" W &372.9" S & 3370.230 & 16.8&
14.0--17.2  & (a) &  \\
29&2005-01a&00:42:28.38&+41:16:36.2 &179.7" W & 27.7" N & 3380.437 & 15.04&
16.2  & (a) &  (8,14) \\
30&2000-08a&00:42:47.47&+41:15:07.5 & 35.5" E & 61.0" S & 1762.733 & 16.96$^e$&
& (g) &  \\
31&2000-08b&00:42:46.76&+41:12:51.8 &27.5" E  &196.7" S & 1752.717 & 17.5$^e$&
& (g) & \\
32&2005-05a&00:42:54.84&+41:16:51.5 &118.5" E & 43.0" N & 3506.566 & 17.2&
13.4  & (b) &  \\
  &        &00:42:54.81&+41:16:51.7 \\
33&2005-05b&00:42:47.15&+41:15:35.7 & 31.9" E & 32.8" S & 3532.699 & 19.5&
& (h) &  \\
  &        &00:42:47.15&+41:15:35.6 \\
34&2005-06a&00:42:28.42&+41:16:50.7 &179.3" W & 42.2" N & 3539.720 & 17.36&
$<51.6 $ & (e) &  \\
35&2005-06b&00:41:37.22&+41:13:11.7 &756.8" W &176.8" S & 3533.729 & 17.79&
& (e) & \\
36&2005-06c&00:42:31.39&+41:16:20.7 &145.8" W & 12.2" N & 3544.710 & 16.52&
6.3  & (i) &  \\
37&2005-07a&00:42:50.71&+41:20:40.2 & 72.0" E &271.7" N & 3581.419 & 17.4&
11.5   & (a) & \\
  &        &00:42:50.79&+41:20:39.9 \\
38&2005-10b&00:42:42.18&+41:18:00.1 & 24.2" W &111.6" N & 3660.316 & 17.6&
$<76.2 $  & (c) & \\
  &        &00:42:42.11&+41:18:00.5 \\
39&2006-02a&00:42:50.68&+41:15:49.1 & 71.7" E & 19.4" S & 3769.248 & 18.0& 
& (c) & \\
  &        &00:42:50.68&+41:15:49.9 \\
40$^f$&2004-11g&00:42:52.62&+41:18:02.4& 93.5" E &  113.9" N & 3315.350 & 18.0
&  &  &(9)  \\
41$^g$&2004-07a&00:42:43.90&+41:17:34.7& 4.8" W & 86.2" N & 3181.517 & 18.3
&  &  &(9)  \\
  &        &00:42:43.89&+41:17:34.9 \\
\noalign{\smallskip}
\hline
\noalign{\smallskip}
\end{tabular}
\label{tab:novae_hornoch_pos}

Notes: \\
\hspace{0.3cm} $^a$: following CBAT nomenclature (see text) \\
\hspace{0.3cm} $^b$: Discoverers of novae:
(a) K. Hornoch, (b) K. Hornoch \& P. Ku\v{s}nir\'ak, (c)
K. Hornoch \& M. Wolf, (d) K. Hornoch \& L. \v{S}arounov\'a, (e) K. Hornoch \& 
D. Mackey, (f) K. Hornoch, P. Garnavich, X. Zhang, T. Pimenova, (g) K. Hornoch
\& D. Carter, (h) K. Hornoch, P. Garnavich, B. Tucker, (i) K. Hornoch \& N. 
Walton\\
\hspace{0.3cm} $^c$: References for detection: 
(1) \citet{2002IAUC.7970....2H}, 
(2) \citet{2003IAUC.8157....4H},
(3) \citet{2003IAUC.8165....2H},
(4) \citet{2003IAUC.8222....2H},
(5) \citet{2003IAUC.8248....3H},
(6) \citet{2004IAUC.8262....2M},
(7) \citet{2004IAUC.8404....2H},
(8) \citet{2005IAUC.8461....1H},
(9) FBR2007\\
References for spectral confirmation:
(10) \citet{2002IAUC.7970....2H},
(11) \citet{2003IAUC.8231....4D},
(12) \citet{2004IAUC.8404....2H},
(13) Filippenko et al. (2004,  see 
http://cfa-www.harvard.edu/iau/CBAT\_M31.html),
(14) \citet{2005IAUC.8462....1D}
\\
\hspace{0.3cm} $^d$: see \citet{2005ASPC..330..449S} \\
\hspace{0.3cm} $^e$: H$_{\alpha}$ magnitudes \\
\hspace{0.3cm} $^f$: discovered in
the WeCAPP program (FBR2007), pre-discovery detections reported her point to an 
earlier date of outburst; there is significantly greater uncertainty in the 
position determination due to the very non-uniform background at the position of
this nova\\
\hspace{0.3cm} $^g$: discovered by Fiaschi et al. (2004, see 
http://cfa-www.harvard.edu/iau/CBAT\_M31.html) and in
the WeCAPP program (FBR2007)\\
\normalsize
\end{table*}

The photometric data (Table~\ref{tab:nova_phot}) are separated by 
novae and are in semi-chronological
order (the first are observations from Lelekovice in chronological
order, then from Ond\v{r}ejov in chronological order and finally
from all other observatories in chronological order). We give the time of 
observation as Julian date JD (Col. 1), brightness  (2) and band used (3).
There are also non detections for which limiting magnitudes are given indicated
by ``[",
mainly closest to the time of discovery. For some novae observations before
discovery are missing because no data were available (nova No. 1)
or because the time between the closest none detection 
before the nova outburst and the nova discovery is quite long (mainly if
nova was found at the begin of observational season of M~31). There are 
mainly R-band (R) plus some V-band (V)
and narrow band H-alpha (Ha) magnitudes.
Measurements with big errors are indicated by a following ":".
Observers, observatory and telescope used are indicated under comment (Col. 4)
and detailed in Table~\ref{tab:hornoch_obs}. 

\begin{table}
\caption{Observers, observatory and telescopes for nova measurements (see
Table~\ref{tab:nova_phot}) }             
\label{tab:hornoch_obs}      
\centering                          
\scriptsize
\begin{tabular}{c c c c}        
\hline\hline                 
ID & Observer & Observatory & Telescope \\   
\hline                        
(1) &  K. Hornoch& Lelekovice &0.35-m \\      
(2) & P. Ku\v{s}nir\'ak, L. \v{S}arounov\'a, M. Wolf& Ond\v{r}ejov &0.65-m \\
(3) & M. Fiaschi& Asiago &1.82-m \\
(4) & P. Ku\v{s}nir\'ak, K. Hornoch &Ond\v{r}ejov &0.65-m \\
(5) & P. Garnavich & VO & 1.83-m  VATT \\
(6) & P. Garnavich & KPNO & 3.5-m  WIYN \\
(7) & P. Garnavich, X. Zhang, T. Pimenova& VO& 1.83-m  VATT \\
(8) & N. Samarasinha, B. Mueller &KPNO &2.1-m \\
(9) & P. Garnavich, B. Tucker & VO & 1.83-m  VATT \\
(10) & D. Mackey &La Palma &2.54-m  INT \\
(11) & P. Garnavich, C. Kennedy &VO &1.83-m  VATT \\
(12) & P. Garnavich, J. Gallagher &VO& 1.83-m  VATT \\
(13) & N. Walton &La Palma &2.54-m  INT \\
(14) & M. Watson &   La Palma& 2.54-m  INT \\
(15) & D. Carter &La Palma& 2.54-m  INT \\
(16) & B. Mueller& KPNO &2.1-m \\
(17) & T. Farnham& KPNO &4-m  Mayall \\
(18) & J. Gallagher, P. Garnavich &KPNO& 0.9-m \\
(19) & M. Burleigh, S. Casewell &La Palma& 2.54-m  INT \\
\hline                                   
\end{tabular}
\normalsize
\end{table}

\begin{table}
\caption{Photometry of novae reported by K. Hornoch. 
Observers, observatory and telescopes for nova measurements are coded in column
comment according to Table~\ref{tab:hornoch_obs}.}
\label{tab:nova_phot}
\scriptsize
\begin{tabular}{llrr}
\hline\hline
JD& Mag & Band & Comment \\
(2\,450\,000+) \\
\hline
\multicolumn{4}{l}{Nova No.1 = M31N~2002-08a}\\
\noalign{\smallskip}
2490.523 &17.05 &R& (1) \\
2504.452 &17.2  &R& (1) \\
2512.360 &18.2  &R& (1) \\
2517.368 &18.6  &R& (1) \\
2520.452 &18.4  &R& (1) \\
2521.433 &18.5  &R& (1) \\
2522.414 &18.7  &R& (1) \\
2525.417 &19.0  &R& (1) \\
2529.435 &18.6  &R& (1) \\
2530.438 &18.7  &R& (1) \\
2547.365 &19.3  &R& (1) \\
2548.505 &19.0  &R& (1) \\
2513.344 &17.83 &R& (2) \\
2516.506 &18.1  &R& (2) \\
2517.560 &18.4  &R& (2) \\
2524.358 &18.5  &R& (2) \\
2530.637 &18.7  &R& (2) \\
\noalign{\smallskip}
\multicolumn{4}{l}{Nova No.2 = M31N~2003-06b}\\
\noalign{\smallskip}
2815.507 &17.2  &R& (1) \\
2816.516 &16.7  &R& (1) \\
2820.522 &18.3  &R& (1) \\
2823.497 &18.6  &R& (1) \\
2828.493 &18.0  &R& (1) \\
2829.511 &18.7  &R& (1) \\
2832.405 &19.0  &R& (1) \\
2835.500 &19.0  &R& (1) \\
2854.514 &18.4  &R& (1) \\
2857.451 &19.2  &R& (1) \\
2858.494 &19.5  &R& (1) \\
2859.400 &19.5  &R& (1) \\
2817.541 &18.2  &R& (2) \\
2819.543 &18.4  &R& (2) \\
2823.534 &18.7  &R& (2) \\
2840.512 &19.5  &R& (2) \\
2856.401 &18.0  &R& (2) \\
2858.583 &19.6  &R& (2) \\
2859.539 &19.8  &R& (2) \\
2863.494 &20.2  &R& (2) \\
2870.00  &19.95 &R& (3) \\
\noalign{\smallskip}
\multicolumn{4}{l}{Nova No.3 = M31N~2003-07b}\\
\noalign{\smallskip}
2831.537& [18.7& R& (1) \\
2835.488& 16.9 & R& (1) \\
2836.487& 17.5 & R& (1) \\
2839.506& 17.5 & R& (1) \\
2840.514& 18.0 & R& (1) \\
2841.495& 18.3 & R& (1) \\
2846.472& 18.6 & R& (1) \\
2847.479& 18.4 & R& (1) \\
2854.531& 19.1 & R& (1) \\
2857.483& 18.8 & R& (1) \\
2858.474& 18.8 & R& (1) \\
2859.490& 19.3 & R& (1) \\
2862.495& 19.0 & R& (1) \\
2867.459& 19.1 & R& (1) \\
2874.479& 19.5 & R& (1) \\
2875.490& 19.3 & R& (1) \\
2876.385& 19.5 & R& (1) \\
2839.534& 17.8 & R& (2) \\
2846.514& 18.8 & R& (2) \\
2847.549& 18.6 & R& (2) \\
2856.408& 18.9 & R& (2) \\
2888.542& 19.7 & R& (2) \\
\noalign{\smallskip}
\multicolumn{4}{l}{Nova No.4 = M31N~2003-06d}\\
\noalign{\smallskip}
2835.488 &[18.8& R& (1) \\
2840.531 &18.5 & R& (1) \\
\noalign{\smallskip}
\multicolumn{4}{l}{Nova No.5 = M31N~2003-09b}\\
\noalign{\smallskip}
2909.312& [18.6& R& (1) \\
2913.335& 17.0 & R& (1) \\
2914.490& 17.2 & R& (1) \\
2925.253& 17.3 & R& (1) \\
2927.340& 17.8 & R& (1) \\
2937.562& 18.1 & R& (1) \\
2946.378& 18.4 & R& (1) \\
2952.573& 18.6 & R& (1) \\
\hline
\end{tabular}
\normalsize
\end{table}
\begin{table}
\addtocounter{table}{-1}
\caption[]{continued. 
	   }
\scriptsize
\begin{tabular}{llrr}
\hline\hline
JD& Mag & Band & Comment \\
(2\,450\,000+) \\
\hline
2973.383& 19.3 & R& (1) \\
2930.476& 18.2 & R& (2) \\
2982.383& 19.9 & R& (2) \\
2983.232& 19.8 & R& (2) \\
\noalign{\smallskip}
\multicolumn{4}{l}{Nova No.6 = M31N~2003-08c}\\
\noalign{\smallskip}
2914.490& [19.0& R& (2) \\
2930.476& 17.9 & R& (4) \\
\noalign{\smallskip}
\multicolumn{4}{l}{Nova No.7 = M31N~2003-11a}\\
\noalign{\smallskip}
2946.378& [18.6& R& (1) \\
2952.573& 16.9 & R& (1) \\
2973.383& 18.7 & R& (1) \\
\noalign{\smallskip}
\multicolumn{4}{l}{Nova No.8 = M31N~2003-12a}\\
\noalign{\smallskip}
2983.232&[19.7 & R& (1) \\
2997.195& 17.8 & R& (1) \\
2997.485& 17.9 & R& (1) \\
2998.325& 19.0 & R& (1) \\
3000.301& 18.8 & R& (1) \\
3010.267& 18.6 & R& (1) \\
3019.240& 19.0 & R& (1) \\
3027.199& 19.4 & R& (1) \\
3028.217& 19.5 & R& (1) \\
3029.235& 19.7 & R& (1) \\
3035.232& 20.0:& R& (1) \\
3028.218& 19.7 & R& (2) \\
3047.300& 20.9:& R& (2) \\
\noalign{\smallskip}
\multicolumn{4}{l}{Nova No.9 = M31N~2004-01a}\\
\noalign{\smallskip}
3019.240& [19.3& R& (1) \\
3027.199& 18.5 & R& (1) \\
3028.217& 18.5 & R& (1) \\
3029.235& 18.2 & R& (1) \\
3035.232& 17.8 & R& (1) \\
3037.336& 18.9 & R& (1) \\
3043.342& 18.1 & R& (1) \\
3047.248& 17.8 & R& (1) \\
3048.220& 18.1 & R& (1) \\
3056.240& 18.5 & R& (1) \\
3060.301& 18.6 & R& (1) \\
3061.238& 18.7 & R& (1) \\
3068.267& 18.9 & R& (1) \\
3070.250& 19.2 & R& (1) \\
3079.272& 19.4 & R& (1) \\
3082.269& 19.6:& R& (1) \\
3028.218& 18.4 & R& (2) \\
3047.300& 17.8 & R& (2) \\
3082.266& 19.9 & R& (2) \\
3289.828& 21.8 & R& (5) \\
3357.568& [22.1& R& (6) \\
\noalign{\smallskip}
\multicolumn{4}{l}{Nova No.10 = M31N~2004-03a}\\
\noalign{\smallskip}
3051.248 &[19.0 &R& (1) \\
3060.301 &18.8  &R& (1) \\
3061.238 &19.0  &R& (1) \\
3068.267 &17.4  &R& (1) \\
3068.301 &17.3  &R& (1) \\
3070.250 &17.8  &R& (1) \\
3079.272 &19.3  &R& (1) \\
3082.269 &19.5: &R& (1) \\
3082.266 &19.5  &R& (2) \\
\noalign{\smallskip}
\multicolumn{4}{l}{Nova No.11 = M31N~2004-03b}\\
\noalign{\smallskip}
3068.267 &[19.8 &R& (1) \\
3070.250 &19.5: &R& (1) \\
3079.272 &18.1  &R& (1) \\
3082.269 &19.5: &R& (1) \\
3082.266 &19.2  &R& (2) \\
\noalign{\smallskip}
\multicolumn{4}{l}{Nova No.12 = M31N~2004-05a}\\
\noalign{\smallskip}
3156.561 &18.2  &R& (1) \\
3164.541 &[19.1 &R& (1) \\
3170.525 &19.0  &R& (1) \\
3178.503 &19.7  &R& (1) \\
3183.501 &20.3: &R& (1) \\
3143.561 &17.8  &R& (2) \\
3144.560 &17.0  &R& (2) \\
3145.563 &17.2  &R& (2) \\
3165.539 &19.2  &R& (2) \\
3171.541 &18.9  &R& (2) \\
3181.517 &20.2  &R& (2) \\
3197.526 &[21.1 &R& (2) \\
\hline
\end{tabular}
\normalsize
\end{table}
\begin{table}
\addtocounter{table}{-1}
\caption[]{continued. 
	   }
\scriptsize
\begin{tabular}{llrr}
\hline\hline
JD& Mag & Band & Comment \\
(2\,450\,000+) \\
\hline
\noalign{\smallskip}
\multicolumn{4}{l}{Nova No.13 = M31N~2004-05b}\\
\noalign{\smallskip}
3156.561 &18.2: &R& (1) \\
3164.541 &18.8  &R& (1) \\
3170.525 &18.8: &R& (1) \\
3143.561 &17.2  &R& (2) \\
3144.560 &17.2  &R& (2) \\
3145.563 &18.1  &R& (2) \\
3165.539 &19.0  &R& (2) \\
3171.541 &19.0: &R& (2) \\
3289.828 &21.7: &R& (5) \\
3357.568 &22.0: &R& (5) \\
\noalign{\smallskip}
\multicolumn{4}{l}{Nova No.14 = M31N~2004-05c}\\
\noalign{\smallskip}
3156.549 &18.9  &R& (1) \\
3164.541 &[19.1 &R& (1) \\
3170.541 &[19.2 &R& (1) \\
3178.521 &[20.2 &R& (1) \\
3183.501 &[20.5 &R& (1) \\
3143.561 &18.8  &R& (2) \\
3144.560 &19.0: &R& (2) \\
3145.563 &18.4  &R& (2) \\
3171.541 &19.3: &R& (2) \\
\noalign{\smallskip}
\multicolumn{4}{l}{Nova No.15 = M31N~2004-06a}\\
\noalign{\smallskip}
3156.561 &[18.9 &R &(1) \\
3164.541 &17.2  &R &(1) \\
3170.525 &17.9  &R &(1) \\
3178.503 &18.8  &R &(1) \\
3183.501 &19.1  &R &(1) \\
3186.505 &19.6: &R &(1) \\
3203.488 &[20.0 &R &(1) \\
3204.483 &[19.4 &R &(1) \\
3165.539 &17.3  &R &(2) \\
3171.541 &17.8  &R &(2) \\
3181.517 &18.9  &R &(2) \\
3197.526 &20.4: &R &(2) \\
\noalign{\smallskip}
\multicolumn{4}{l}{Nova No.16 = M31N~2004-06b}\\
\noalign{\smallskip}
3164.540 &[19.4 &R &(1) \\
3170.525 &18.6  &R &(1) \\
3178.503 &17.9  &R &(1) \\
3183.501 &18.0  &R &(1) \\
3186.505 &18.0  &R &(1) \\
3193.506 &18.1  &R &(1) \\
3203.488 &18.6  &R &(1) \\
3204.483 &18.3  &R &(1) \\
3206.458 &18.3  &R &(1) \\
3206.476 &18.5  &R &(1) \\
3208.509 &18.7  &R &(1) \\
3212.452 &18.7  &R &(1) \\
3217.470 &18.9  &R &(1) \\
3222.901 &19.2  &R &(1) \\
3225.405 &19.2  &R &(1) \\
3227.505 &19.4  &R &(1) \\
3171.541 &18.1  &R &(2) \\
3181.517 &17.8  &R &(2) \\
3197.526 &18.3  &R &(2) \\
3209.399 &18.7  &R &(2) \\
3220.474 &19.2  &R &(2) \\
3224.497 &19.4  &R &(2) \\
3225.482 &19.3  &R &(2) \\
3226.570 &19.8: &R &(2) \\
3229.557 &19.9: &R &(2) \\
3236.586 &19.9: &R &(2) \\
3289.828 &[22.0 &R &(5) \\
\noalign{\smallskip}
\multicolumn{4}{l}{Nova No.17 = M31N~2004-06c}\\
\noalign{\smallskip}
3164.540 &[19.4 &R &(1) \\
3178.521 &18.0  &R &(1) \\
3183.526 &17.4  &R &(1) \\
3186.505 &18.4  &R &(1) \\
3193.506 &19.3  &R &(1) \\
3203.459 &[19.0 &R &(1) \\
3203.488 &[19.3 &R &(1) \\
3181.517 &17.1  &R &(2) \\
3197.526 &19.3  &R &(2) \\
3289.828 &[22.0 &R &(5) \\
\hline
\end{tabular}
\normalsize
\end{table}
\begin{table}
\addtocounter{table}{-1}
\caption[]{continued. 
	   }
\scriptsize
\begin{tabular}{llrr}
\hline\hline
JD& Mag & Band & Comment \\
(2\,450\,000+) \\
\hline
\noalign{\smallskip}
\multicolumn{4}{l}{Nova No.18 = M31N~2003-07a}\\
\noalign{\smallskip}
2829.525 &[18.7 &R &(1) \\
2831.538 &18.3: &R &(1) \\
2835.489 &18.7: &R &(1) \\
2840.514 &18.4  &R &(1) \\
2846.474 &17.9  &R &(1) \\
2847.480 &17.6  &R &(1) \\
2857.481 &18.7  &R &(1) \\
2859.489 &18.8  &R &(1) \\
2867.459 &18.7: &R &(1) \\
2874.480 &19.2: &R &(1) \\
2875.471 &18.7  &R &(1) \\
2878.430 &18.9: &R &(1) \\
2897.328 &18.5  &R &(1) \\
2904.322 &18.6: &R &(1) \\
2913.325 &19.1  &R &(1) \\
3203.488 &20.0  &R &(1) \\
2530.636 &[20.6 &R &(2) \\
2839.534 &19.0  &R &(2) \\
2846.520 &18.2  &R &(2) \\
2847.549 &18.0  &R &(2) \\
2850.542 &18.2  &R &(2) \\
2856.409 &18.7  &R &(2) \\
2888.543 &18.7  &R &(2) \\
\noalign{\smallskip}
\multicolumn{4}{l}{Nova No.19 = M31N~2004-08a}\\
\noalign{\smallskip}
3217.470 &[19.0 &R &(1) \\
3221.423 &17.7  &R &(1) \\
3222.4   &18.3  &R &(1) \\
3220.474 &17.4  &R &(2) \\
3224.497 &19.0: &R &(2) \\
3226.570 &[19.3 &R &(2) \\
\noalign{\smallskip}
\multicolumn{4}{l}{Nova No.20 = M31N~2004-08b}\\
\noalign{\smallskip}
3221.460 &19.2: &R &(1) \\
3222.401 &18.8  &R &(1) \\
3225.405 &17.4  &R &(1) \\
3227.505 &18.3  &R &(1) \\
3228.393 &18.0  &R &(1) \\
3228.456 &17.7  &R &(1) \\
3233.443 &17.6  &R &(1) \\
3235.370 &17.6  &R &(1) \\
3236.410 &17.7  &R &(1) \\
3237.375 &17.7  &R &(1) \\
3240.429 &17.6  &R &(1) \\
3241.402 &17.5  &R &(1) \\
3246.342 &18.3  &R &(1) \\
3246.399 &18.1  &R &(1) \\
3249.415 &18.6  &R &(1) \\
3249.464 &18.5  &R &(1) \\
3252.312 &18.7  &R &(1) \\
3252.368 &18.7  &R &(1) \\
3253.319 &19.0  &R &(1) \\
3254.362 &19.0  &R &(1) \\
3255.430 &19.0  &R &(1) \\
3257.429 &18.3  &R &(1) \\
3257.453 &18.4  &R &(1) \\
3258.363 &18.4  &R &(1) \\
3258.393 &18.3  &R &(1) \\
3259.384 &18.6  &R &(1) \\
3259.413 &18.4  &R &(1) \\
3262.453 &18.6  &R &(1) \\
3265.332 &19.4  &R &(1) \\
3265.352 &19.5  &R &(1) \\
3266.335 &19.4  &R &(1) \\
3270.330 &19.6: &R &(1) \\
3275.290 &19.6  &R &(1) \\
3282.281 &19.4  &R &(1) \\
3288.272 &19.2  &R &(1) \\
3220.474 &[20.5 &R &(2) \\
3224.497 &17.8  &R &(2) \\
3225.482 &17.3  &R &(2) \\
3226.570 &17.8  &R &(2) \\
3229.557 &18.3  &R &(2) \\
3236.586 &17.7  &R &(2) \\
3241.560 &17.6  &R &(2) \\
3241.563 &17.6  &V &(2) \\
3251.518 &18.6  &R &(2) \\
3253.558 &19.0  &R &(2) \\
3257.571 &18.5  &R &(2) \\
3260.385 &18.6  &R &(2) \\
3279.356 &19.6  &R &(2) \\
3283.619 &19.5  &R &(2) \\
3301.408 &20.2: &R &(2) \\
3289.828 &20.40 &R &(5) \\
\hline
\end{tabular}
\normalsize
\end{table}
\begin{table}
\addtocounter{table}{-1}
\caption[]{continued. 
	   }
\scriptsize
\begin{tabular}{llrr}
\hline\hline
JD& Mag & Band & Comment \\
(2\,450\,000+) \\
\hline
\noalign{\smallskip}
\multicolumn{4}{l}{Nova No.21 = M31N~2004-09a}\\
\noalign{\smallskip}
3246.399 &[19.1 &R &(1) \\
3249.415 &18.3  &R &(1) \\
3249.438 &18.0  &R &(1) \\
3249.452 &17.8  &R &(1) \\
3252.312 &17.7  &R &(1) \\
3252.368 &17.6  &R &(1) \\
3253.319 &18.3  &R &(1) \\
3253.341 &18.2  &R &(1) \\
3253.441 &18.0  &R &(1) \\
3254.362 &18.1  &R &(1) \\
3254.400 &18.2  &R &(1) \\
3255.430 &18.0  &R &(1) \\
3257.429 &18.2  &R &(1) \\
3257.453 &18.1  &R &(1) \\
3258.363 &18.3  &R &(1) \\
3258.393 &18.4  &R &(1) \\
3259.384 &18.6  &R &(1) \\
3259.413 &18.4  &R &(1) \\
3262.453 &18.7  &R &(1) \\
3262.472 &18.6  &R &(1) \\
3265.332 &18.8  &R &(1) \\
3265.352 &18.5: &R &(1) \\
3266.335 &19.2: &R &(1) \\
3270.330 &18.6: &R &(1) \\
3270.351 &19.1: &R &(1) \\
3275.290 &19.5: &R &(1) \\
3279.398 &19.4  &R &(1) \\
3282.257 &19.2  &R &(1) \\
3288.272 &19.6: &R &(1) \\
3288.292 &20.1: &R &(1) \\
3241.560 &[19.8 &R &(2) \\
3251.518 &17.5  &R &(2) \\
3253.558 &18.1  &R &(2) \\
3257.571 &18.2  &R &(2) \\
3260.385 &18.6  &R &(2) \\
3279.356 &19.3  &R &(2) \\
3283.619 &19.8: &R &(2) \\
3289.802 &20.8: &R &(5) \\
\noalign{\smallskip}
\multicolumn{4}{l}{Nova No.22 = M31N~2004-10a}\\
\noalign{\smallskip}
3301.408 &18.9: &R &(2) \\
3287.835 &18.64 &R &(7) \\
3288.802 &18.60 &R &(7) \\
3357.568 &19.6  &R &(6) \\
3405.633 &[20.5 &R &(8) \\
\noalign{\smallskip}
\multicolumn{4}{l}{Nova No.23 = M31N~2004-10b}\\
\noalign{\smallskip}
3301.408 &18.6: &R &(2) \\
3287.835 &19.63 &R &(7) \\
3288.802 &19.06 &R &(7) \\
3357.568 &[21.4 &R &(6) \\
\noalign{\smallskip}
\multicolumn{4}{l}{Nova No.24 = M31N~2004-11b}\\
\noalign{\smallskip}
3315.347 &16.7  &R &(1) \\
3315.390 &16.6  &R &(1) \\
3317.352 &17.1  &R &(1) \\
3321.404 &17.4  &R &(1) \\
3324.305 &17.3  &R &(1) \\
3325.218 &17.2  &R &(1) \\
3334.218 &17.5  &R &(1) \\
3335.273 &17.6  &R &(1) \\
3339.296 &17.9  &R &(1) \\
3339.318 &17.9  &R &(1) \\
3344.192 &18.1  &R &(1) \\
3344.214 &18.2  &R &(1) \\
3346.410 &18.3  &R &(1) \\
3347.344 &18.5  &R &(1) \\
3347.370 &18.8  &R &(1) \\
3348.359 &18.7  &R &(1) \\
3360.236 &19.3  &R &(1) \\
3361.324 &19.1  &R &(1) \\
3370.230 &19.5  &R &(1) \\
3370.267 &19.3  &R &(1) \\
3378.391 &19.7  &R &(1) \\
3380.426 &19.7  &R &(1) \\
3381.276 &19.4  &R &(1) \\
3381.417 &19.4  &R &(1) \\
3382.219 &19.3  &R &(1) \\
3384.212 &19.6: &R &(1) \\
3387.400 &20.0: &R &(1) \\
3301.408 &[21   &R &(2) \\
3342.172 &18.2  &R &(2) \\
3358.260 &19.2  &R &(2) \\
3373.321 &19.4  &R &(2) \\
\hline
\end{tabular}
\normalsize
\end{table}
\begin{table}
\addtocounter{table}{-1}
\caption[]{continued. 
	   }
\scriptsize
\begin{tabular}{llrr}
\hline\hline
JD& Mag & Band & Comment \\
(2\,450\,000+) \\
\hline
3377.266 &19.5  &R &(2) \\
3381.249 &19.5  &R &(2) \\
3381.253 &19.5: &V &(2) \\
3382.246 &19.3: &R &(2) \\
3387.231 &19.8: &R &(2) \\
3388.224 &19.5: &R &(2) \\
3357.568 &19.1  &R &(6) \\
\noalign{\smallskip}
\multicolumn{4}{l}{Nova No.25 = M31N~2004-11a}\\
\noalign{\smallskip}
3315.347 &16.6  &R &(1) \\
3315.390 &16.5  &R &(1) \\
3317.352 &17.0  &R &(1) \\
3321.404 &18.1  &R &(1) \\
3324.305 &18.2  &R &(1) \\
3325.218 &18.4  &R &(1) \\
3334.218 &19.0  &R &(1) \\
3335.273 &19.2  &R &(1) \\
3339.296 &19.4: &R &(1) \\
3339.318 &19.6: &R &(1) \\
3344.192 &19.9: &R &(1) \\
3344.214 &19.3: &R &(1) \\
3301.408 &[21   &R &(2) \\
3342.192 &19.3: &R &(2) \\
3358.260 &20.5: &R &(2) \\
3357.568 &19.8  &R &(6) \\
\noalign{\smallskip}
\multicolumn{4}{l}{Nova No.26 = M31N~2004-11d}\\
\noalign{\smallskip}
3325.218 &[18.5:&R &(1) \\
3334.218 &17.0: &R &(1) \\
3335.273 &17.3: &R &(1) \\
3339.296 &17.6: &R &(1) \\
3344.192 &18.8: &R &(1) \\
3342.172 &18.5: &R &(2) \\
3357.568 &19.4  &R &(6) \\
3405.633 &[20.5 &R &(8) \\
\noalign{\smallskip}
\multicolumn{4}{l}{Nova No.27 = M31N~2004-11e}\\
\noalign{\smallskip}
3321.417 &[19.5 &R &(1) \\
3339.296 &17.6  &R &(1) \\
3339.339 &17.6  &R &(1) \\
3344.192 &18.5  &R &(1) \\
3346.410 &18.8  &R &(1) \\
3347.344 &18.7  &R &(1) \\
3348.359 &19.0  &R &(1) \\
3360.236 &19.5  &R &(1) \\
3361.356 &19.4  &R &(1) \\
3370.230 &19.5  &R &(1) \\
3378.391 &19.7  &R &(1) \\
3380.426 &19.7  &R &(1) \\
3381.276 &20.0: &R &(1) \\
3381.417 &19.9: &R &(1) \\
3358.260 &19.2  &R &(2) \\
3377.266 &19.7  &R &(2) \\
3288.848 &[22   &R &(7) \\
\noalign{\smallskip}
\multicolumn{4}{l}{Nova No.28 = M31N~2004-12a}\\
\noalign{\smallskip}
3361.356 &19.2  &R &(1) \\
3370.230 &16.8  &R &(1) \\
3370.267 &16.9  &R &(1) \\
3377.293 &18.1  &R &(1) \\
3378.391 &18.1  &R &(1) \\
3380.224 &18.3  &R &(1) \\
3380.426 &18.4  &R &(1) \\
3381.276 &18.5  &R &(1) \\
3381.417 &18.5  &R &(1) \\
3382.219 &18.4  &R &(1) \\
3382.246 &18.4  &R &(1) \\
3384.212 &18.8  &R &(1) \\
3386.202 &18.6  &R &(1) \\
3386.424 &18.8  &R &(1) \\
3387.400 &18.8  &R &(1) \\
3394.234 &19.4: &R &(1) \\
3407.231 &19.7: &R &(1) \\
3373.321 &17.4  &R &(2) \\
3377.266 &18.0  &R &(2) \\
3381.249 &18.4  &R &(2) \\
3381.253 &18.9: &V &(2) \\
3382.246 &18.3  &R &(2) \\
3387.231 &18.7  &R &(2) \\
3388.224 &18.8  &R &(2) \\
3406.326 &19.7  &R &(2) \\
3411.257 &19.7  &R &(2) \\
3357.568 &[22   &R &(6) \\
\hline
\end{tabular}
\normalsize
\end{table}
\begin{table}
\addtocounter{table}{-1}
\caption[]{continued. 
	   }
\scriptsize
\begin{tabular}{llrr}
\hline\hline
JD& Mag & Band & Comment \\
(2\,450\,000+) \\
\hline
\noalign{\smallskip}
\multicolumn{4}{l}{Nova No.29 = M31N~2005-01a}\\
\noalign{\smallskip}
3377.293 &19.4  &R &(1) \\
3378.391 &17.9  &R &(1) \\
3380.224 &15.72 &R &(1) \\
3380.253 &15.68 &R &(1) \\
3380.437 &15.33 &R &(1) \\
3381.266 &15.32 &R &(1) \\
3381.286 &15.27 &R &(1) \\
3381.306 &15.30 &R &(1) \\
3381.408 &15.34 &R &(1) \\
3381.426 &15.31 &R &(1) \\
3382.219 &15.16 &R &(1) \\
3382.257 &15.21 &R &(1) \\
3384.212 &15.05 &R &(1) \\
3384.252 &15.10 &R &(1) \\
3384.315 &14.95 &I &(1) \\
3384.405 &15.04 &R &(1) \\
3386.202 &15.27 &R &(1) \\
3386.228 &15.26 &R &(1) \\
3386.424 &15.31 &R &(1) \\
3387.212 &15.36 &R &(1) \\
3387.249 &15.34 &R &(1) \\
3387.400 &15.34 &R &(1) \\
3390.287 &15.52 &R &(1) \\
3390.327 &15.50 &R &(1) \\
3394.234 &15.80 &R &(1) \\
3398.274 &16.61 &R &(1) \\
3401.212 &17.17 &R &(1) \\
3407.231 &[19.9 &R &(1) \\
3407.283 &[19.8 &R &(1) \\
3373.321 &[19.8 &R &(2) \\
3377.266 &19.2  &R &(2) \\
3381.249 &15.48 &R &(2) \\
3381.253 &15.45 &V &(2) \\
3382.246 &15.21 &R &(2) \\
3382.251 &15.26 &V &(2) \\
3384.326 &15.12 &R &(2) \\
3384.333 &15.26 &V &(2) \\
3386.300 &15.27 &R &(2) \\
3386.305 &15.65 &V &(2) \\
3387.231 &15.39 &R &(2) \\
3387.237 &15.75 &V &(2) \\
3388.224 &15.42 &R &(2) \\
3388.233 &15.77 &V &(2) \\
3405.273 &19.4: &R &(2) \\
3406.326 &20.1  &R &(2) \\
3411.257 &[20.3 &R &(2) \\
3257.568 &[21   &R &(6) \\
3405.633 &19.7  &R &(8) \\
3509.952 &21.7  &R &(9) \\
3532.699 &21.13 &R &(10) \\
3534.701 &21.38 &R &(10) \\
3535.681 &21.45 &R &(10) \\
3538.697 &21.43 &R &(10) \\
3541.718 &21.2  &R &(10) \\
3651.821 &22.0: &R &(11) \\
3702.638 &[21.3 &R &(5) \\
3710.730 &21.9  &R &(12) \\
3532.711 &19.09 &Ha&(10) \\
3534.696 &19.03 &Ha&(10) \\
3535.695 &18.98 &Ha&(10) \\
3544.705 &18.94 &Ha&(13)\\
\noalign{\smallskip}
\multicolumn{4}{l}{Nova No.30 = M31N~2000-08a}\\
\noalign{\smallskip}
1752.717 &18.7: &Ha&(14) \\
1762.733 &17.09 &Ha&(15) \\
1762.741 &16.96 &Ha&(15) \\
3357.568 &[22   &R &(6) \\
\noalign{\smallskip}
\multicolumn{4}{l}{Nova No.31 = M31N~2000-08b}\\
\noalign{\smallskip}
1752.717 &17.5  &Ha&(14) \\
1762.733 &17.91 &Ha&(15) \\
1762.741 &17.90 &Ha&(15) \\
3357.568 &[22   &R &(6) \\
\hline
\end{tabular}
\normalsize
\end{table}
\begin{table}
\addtocounter{table}{-1}
\caption[]{continued. 
	   }
\scriptsize
\begin{tabular}{llrr}
\hline\hline
JD& Mag & Band & Comment \\
(2\,450\,000+) \\
\hline
\noalign{\smallskip}
\multicolumn{4}{l}{Nova No.32 = M31N~2005-05a}\\
\noalign{\smallskip}
3373.321 &[19.8 &R &(2) \\
3450.265 &[19.0:&R &(2) \\
3506.566 &17.2  &R &(2) \\
3510.542 &18.4  &R &(2) \\
3511.539 &18.4  &R &(2) \\
3516.561 &18.7  &R &(2) \\
3518.554 &19.1  &R &(2) \\
3508.961 &18.42 &R &(9) \\
3509.952 &18.35 &R &(9) \\
3527.927 &20.1  &R &(16) \\
3532.699 &21.4  &R &(10) \\
3532.711 &18.50 &Ha&(10) \\
3534.696 &18.90 &Ha&(10) \\
3535.681 &22.0  &R &(10) \\
3535.695 &19.10 &Ha&(10) \\
3538.697 &22.1  &R &(10) \\
3541.718 &[21.8 &R &(10) \\
\noalign{\smallskip}
\multicolumn{4}{l}{Nova No.33 = M31N~2005-05b}\\
\noalign{\smallskip}
3301.068 &[22   &R &(5) \\
3405.672 &[21.5 &R &(8) \\
3508.961 &20.2  &R &(9) \\
3509.952 &20.0  &R &(9) \\
3527.943 &20.5  &R &(16) \\
3532.699 &19.5  &R &(10) \\
3532.711 &17.73 &Ha&(10) \\
3534.696 &17.78 &Ha&(10) \\
3534.701 &19.8  &R &(10) \\
3535.681 &19.9  &R &(10) \\
3535.695 &17.84 &Ha&(10) \\
3538.697 &19.9  &R &(10) \\
3539.720 &19.9  &R &(10) \\
3541.718 &20.0  &R &(10) \\
3544.705 &17.51 &Ha&(13)\\
3544.710 &20.0  &R &(13)\\
3545.698 &19.9  &R &(13)\\
3651.821 &[21.7 &R &(11) \\
3702.638 &[21.3 &R &(11) \\
\noalign{\smallskip}
\multicolumn{4}{l}{Nova No.34 = M31N~2005-06a}\\
\noalign{\smallskip}
3550.554 &18.8  &R &(2) \\
3564.493 &19.0  &R &(2) \\
3301.068 &[22   &R &(5) \\
3509.952 &[21.6 &R &(9) \\
3527.943 &[20.5 &R &(16) \\
3532.699 &[22.0 &R &(10) \\
3533.729 &21.1  &R &(10) \\
3534.696 &18.55 &Ha&(10) \\
3534.701 &19.07 &R &(10) \\
3535.681 &18.08 &R &(10) \\
3535.695 &17.56 &Ha&(10) \\
3538.697 &17.63 &R &(10) \\
3539.720 &17.36 &R &(10) \\
3541.718 &18.47 &R &(10) \\
3544.705 &16.36 &Ha&(13)\\
3544.710 &18.14 &R &(13)\\
3545.698 &18.18 &R &(13)\\
3553.847 &18.10 &R &(17) \\
3651.821 &[21.7 &R &(11) \\
3702.638 &[21.3 &R &(11) \\
\noalign{\smallskip}
\multicolumn{4}{l}{Nova No.35 = M31N~2005-06b}\\
\noalign{\smallskip}
3257.568 &[21.5 &R &(6) \\
3532.699 &17.97 &R &(10) \\
3532.711 &17.66 &Ha&(10) \\
3533.729 &17.79 &R &(10) \\
3534.696 &17.70 &Ha&(10) \\
3534.701 &18.00 &R &(10) \\
3535.681 &17.97 &R &(10) \\
3535.695 &17.34 &Ha&(10) \\
3538.697 &18.19 &R &(10) \\
3539.720 &18.21 &R &(10) \\
3541.718 &18.43 &R &(10) \\
3544.705 &16.86 &Ha&(13)\\
3544.710 &18.82 &R &(13)\\
3545.698 &18.89 &R &(13)\\
3553.847 &19.33 &R &(17) \\
\noalign{\smallskip}
\multicolumn{4}{l}{Nova No.36 = M31N~2005-06c}\\
\noalign{\smallskip}
3550.554 &18.4  &R &(2) \\
3301.068 &[22   &R &(5) \\
3509.952 &[21.6 &R &(9) \\
\hline
\end{tabular}
\normalsize
\end{table}
\begin{table}
\addtocounter{table}{-1}
\caption[]{continued. 
	   }
\scriptsize
\begin{tabular}{llrr}
\hline\hline
JD& Mag & Band & Comment \\
(2\,450\,000+) \\
\hline
3527.943 &[20.5 &R &(16) \\
3538.697 &[22.0 &R &(10) \\
3541.718 &[21.6 &R &(10) \\
3544.705 &15.66 &Ha&(13)\\
3544.710 &16.52 &R &(13)\\
3545.698 &17.12 &R &(13)\\
3553.847 &19.75 &R &(17) \\
\noalign{\smallskip}
\multicolumn{4}{l}{Nova No.37 = M31N~2005-07a}\\
\noalign{\smallskip}
3575.429 &[19.0 &R &(1) \\
3579.409 &18.4  &R &(1) \\
3581.419 &17.4  &R &(1) \\
3584.404 &19.1  &R &(1) \\
3588.415 &18.8  &R &(1) \\
3564.493 &[19.5 &R &(2) \\
3587.509 &19.3  &R &(2) \\
3594.390 &19.7  &R &(2) \\
3553.847 &[21.7 &R &(17) \\
3651.862 &19.53 &R &(11) \\
3702.638 &20.8  &R &(5) \\
3710.730 &21.03 &R &(12) \\
3760.586 &21.8  &R &(18) \\
3771.346 &21.7  &R &(19) \\
\noalign{\smallskip}
\multicolumn{4}{l}{Nova No.38 = M31N~2005-10b}\\
\noalign{\smallskip}
3658.303 & [18.8&R &(1) \\
3660.316 & 17.6 &R &(1) \\
3663.235 & 17.9 &R &(1) \\
3670.281 & 17.9 &R &(1) \\
3671.229 & 18.2 &R &(1) \\
3674.237 & 18.5 &R &(1) \\
3676.221 & 18.1 &R &(1) \\
3683.218 & 18.6 &R &(1) \\
3730.266 & 19.5 &R &(1) \\
3670.324 & 18.2 &R &(2) \\
3671.256 & 18.3 &R &(2) \\
3673.238 & 18.1 &R &(2) \\
3678.317 & 18.6 &R &(2) \\
3686.565 & 18.4 &R &(2) \\
3651.862 &[21   &R &(11) \\
3702.638 &18.95 &R &(5) \\
3710.730 &19.31 &R &(12) \\
3760.586 &20.1  &R &(18) \\
3771.346 &19.9  &R &(19) \\
\noalign{\smallskip}
\multicolumn{4}{l}{Nova No.39 = M31N~2006-02a}\\
\noalign{\smallskip}
3765.234 & [18.8&R &(1) \\
3771.264 & 19.2 &R &(1) \\
3769.248 & 18.0 &R &(2) \\
3710.727 & [21  &R &(12) \\
3771.346 & 19.03&R &(19) \\
\noalign{\smallskip}
\multicolumn{4}{l}{Nova No.40 = M31N~2004-11g}\\
\noalign{\smallskip}
3301.409 &[20.0& R& (2) \\
3315.347 & 18.1& R& (1) \\
3315.390 & 18.0& R& (1) \\
3317.352 & 18.4& R& (1) \\
3321.404 & 18.4& R& (1) \\
\noalign{\smallskip}
\multicolumn{4}{l}{Nova No.41 = M31N~2004-07a}\\
\noalign{\smallskip}
3143.561 & 19.5:& R & (2) \\
3145.563 & 19.1:& R & (2) \\
3171.541 & 19.5:& R & (2) \\
3181.517 & 18.3 & R & (2) \\
3197.526 & 18.3 & R & (2) \\
3209.399 & 18.4 & R & (2) \\
3220.474 & 18.7 & R & (2) \\
3224.497 & 19.0 & R & (2) \\
3225.482 & 18.6 & R & (2) \\
3226.570 & 19.4 & R & (2) \\
3229.557 & 19.1 & R & (2) \\
3236.586 & 20.0:& R & (2) \\
3241.560 & 18.9:& R & (2) \\
3253.558 & 19.9:& R & (2) \\
3301.408 & 19.8:& R & (2) \\
3342.192 & 19.2:& R & (2) \\
3358.260 & 19.4:& R & (2) \\
3532.711 & 18.02& Ha& (10) \\
3534.696 & 18.07& Ha& (10) \\
3535.695 & 17.9 & Ha& (10) \\
3544.705 & 18.5 & Ha& (13) \\
\hline
\end{tabular}
\normalsize
\end{table}

\end{document}